\documentclass[11pt]{revtex4-1}
\usepackage{amsmath,amssymb,graphicx,braket,hyperref}
\usepackage{multirow}
\usepackage{float}
\usepackage{bm}

\begin{document}

\title{$\mathbb{Z}_N$ stability and continuity in twisted Eguchi-Kawai model \\with two-flavor adjoint fermions}

\author{Yudai Hamada}
\email{2433310108h@kindai.ac.jp}
\address{Department of Physics, Kindai University,  Osaka 577-8502, Japan}

\author{Tatsuhiro Misumi}
\email{misumi@phys.kindai.ac.jp}
\address{Department of Physics, Kindai University,  Osaka 577-8502, Japan}

\begin{abstract}
We investigate the twisted Eguchi-Kawai (TEK) reduced model of four-dimensional $SU(N)$ gauge theory in the presence of two-flavor adjoint fermions (adjoint TEK model). Using Monte Carlo simulations with $N=121$, twist parameter $k=1$, hopping parameter $\kappa=0.01$-$0.03$ ($\ll\kappa_c $) and inverse 't Hooft coupling $b=0.30$-$0.45$, we show that heavy adjoint fermions stabilize the $(\mathbb{Z}_N)^4$ center-symmetric vacuum even for the minimal twist satisfying $k/\sqrt{N} < 1/9$, where the $(\mathbb{Z}_N)^4$ symmetry is spontaneously broken in the absence of adjoint fermions. 
This result also suggests that the adjoint TEK model with the minimal twist is equivalent to $SU(N)$ gauge theory over a broader $(\kappa,b)$ parameter region than the adjoint EK model without twist. 
We further extend our analysis to a partially reduced model to realize a geometry akin to $\mathbb{R}^3 \times S^1$ and study the theory under $S^1$ compactification with periodic adjoint fermions.
Numerical simulations with $N=16$-$49$, $b=0.30$-$0.46$ and $\kappa=0.03$-$0.16$ are consistent with the adiabatic continuity conjecture: with periodic adjoint fermions, the theory remains in a center-symmetric (confined) phase as the $S^1$ circle size is reduced, in contrast to the deconfining transition observed in the pure TEK model or in the TEK model with antiperiodic adjoint fermions. We present the Polyakov loop measurements and consistency checks supporting these findings.
\end{abstract}

\maketitle 

\newpage
\tableofcontents
\newpage
 
\section{Introduction}
Large-$N$ reduction \cite{tHooft:1973alw,Eguchi:1982nm} posits that in the limit of $N\to\infty$, $SU(N)$ gauge theory becomes independent of volume, allowing one to replace the infinite-volume lattice with a much smaller system, even a single-site lattice model, while retaining the physics of the full theory (see \cite{Lucini:2001ej,Lucini:2012gg} for review). The simplest one-site model -- now known as the Eguchi-Kawai (EK) model \cite{Eguchi:1982nm} -- contains $SU(N)$ link variables $U_\mu$ on one site with the standard Wilson action,
\begin{equation}
S_{\rm EK} \,=\, -bN \sum_{\mu\neq\nu=1}^4 \mathrm{Tr}\Big( U_\mu U_\nu U_\mu^\dagger U_\nu^\dagger \Big)\,,
\label{eq:EK0}
\end{equation} 
with $b=1/(g^2 N)$ related to $\beta$ in lattice QCD.
Volume independence requires preservation of the $(\mathbb{Z}_N)^4$ center symmetry, under which each link transforms as $U_\mu \to z_\mu U_\mu$ with $z_\mu\in\mathbb{Z}_N$. If the center symmetry remains unbroken for all four directions as $\langle \mathrm{Tr} U_\mu\rangle = 0$, then large-$N$ expectation values of Wilson loops in the one-site model coincide with those on the 4d infinite lattice. However, it was quickly realized \cite{Bhanot:1982sh} that the naive EK model fails in the weak-coupling regime: the $(\mathbb{Z}_N)^4$ center symmetry breaks spontaneously, typically signaled by $\langle \mathrm{Tr}U_\mu\rangle \neq 0$ for all $\mu$ at sufficiently large $b$. This breakdown invalidates volume reduction in that regime, matching the observation that the one-site model undergoes a phase transition to a $\mathbb{Z}_N$-broken phase at large $b$ \cite{Bhanot:1982sh}. 

Several methods have been proposed to preserve the center symmetry in reduced models \cite{Bhanot:1982sh,Gonzalez-Arroyo:1982hwr, Gonzalez-Arroyo:1982hyq, Narayanan:2003fc, Bringoltz:2009kb, Bringoltz:2011by}. 
Among them, the \emph{twisted} Eguchi-Kawai (TEK) model introduced by Gonzalez-Arroyo and Okawa \cite{Gonzalez-Arroyo:1982hwr,Gonzalez-Arroyo:1982hyq}, modifies the EK action by including a constant `twist' phase $z_{\mu\nu}\in \mathbb{Z}_N$ in the plaquettes. The TEK action is
\begin{equation}
S_{\rm TEK} \,=\, -bN \sum_{\mu\neq\nu=1}^4 \mathrm{Tr}\Big( z_{\mu\nu} U_\mu U_\nu U_\mu^\dagger U_\nu^\dagger \Big)\,,
\label{eq:TEK0}
\end{equation}
where $z_{\mu\nu}=\exp(2\pi i k/L) = z_{\nu\mu}^{*}$ are $\mathbb{Z}_N$ phases, with integers $k$, $L$ such that $N=L^2$ \cite{Gonzalez-Arroyo:2010omx}. The twist induces a background flux that geometrically frustrates the tendency of the Polyakov lines to align, thus helping to preserve $\mathbb{Z}_N$ symmetry. Indeed, for appropriate choices of $(k,L)$ (typically $k$ and $L$ coprime, and $k/L$ held fixed as $N\to\infty$), the TEK model was shown to maintain $(\mathbb{Z}_N)^4$ center symmetry and reproduce $SU(N)$ Yang-Mills and QCD results in large $N$, including string tension \cite{Gonzalez-Arroyo:2012euf}, meson spectrum \cite{Gonzalez-Arroyo:2015bya,Perez:2020vbn} and chiral condensate \cite{Bonanno:2023ypf}. A drawback is that one must tune the twist parameter $k$ as $N\to \infty$ to stay on the center-symmetric branch (at least need to keep $k/L > 1/9$) \cite{Gonzalez-Arroyo:2010omx}. 

Another approach is to introduce adjoint fermion fields in the reduced model. In the original adjoint Eguchi-Kawai (AEK) model, one adds $N_f$ flavors of Wilson-Dirac fermions in the adjoint representation to the one-site EK action \cite{Bringoltz:2009kb,Bringoltz:2011by}
\begin{equation}
S_{\rm AEK} \,=\, -bN \sum_{\mu\neq\nu = 1}^4 \mathrm{Tr}(U_\mu U_\nu U_\mu^\dagger U_\nu^\dagger)+\sum_{f=1}^{N_f} \bar\psi_f D_W[U]\psi_f\,,
\label{eq:AEK}
\end{equation}
where $D_W$ is the Wilson-Dirac operator on the one-site lattice, including a hopping parameter $\kappa$ and adjoint link variables $U_\mu^{\rm adj}$. The adjoint fermions do not transform under the center (since the $\mathbb{Z}_N$ phases cancel out in the adjoint representation), so their presence can exert an effective potential on the Polyakov loops that favors the $(\mathbb{Z}_N)^4$ center-symmetric vacuum. In fact, the simulation studies by Bringoltz and Sharpe \cite{Bringoltz:2009kb, Bringoltz:2011by}, showed that a single adjoint flavor can help stabilize the $(\mathbb{Z}_N)^4$ symmetry in a certain region of ($\kappa,b$). However, a purely adjoint-fermion based reduced model without twist has its own limitations: when all four directions are collapsed to a point, one cannot easily extract physical observables like meson masses because there is no notion of spatial separation (no ``emergent" space). Moreover, for the weak-coupling (large $b$) and heavy-mass region, the $(\mathbb{Z}_N)^4$ center symmetry is expected to be broken, thus one needs to tune $\kappa$ 
near the chiral limit $\kappa_c$ to maintain the $(\mathbb{Z}_N)^4$ center symmetry \cite{Bringoltz:2009kb,Bringoltz:2011by}. Recently, it has been also argued that this model contains relatively large $1/N$ corrections \cite{Gonzalez-Arroyo:2012ztz, Gonzalez-Arroyo:2013bta, GarciaPerez:2013dgk, GarciaPerez:2015rda}.
This is where combining both ideas -- adjoint fermions and a twist -- becomes advantageous.

In this work, we study the Twisted Eguchi-Kawai model with two flavors of adjoint Wilson fermions (adjoint TEK model) \cite{Azeyanagi:2010ne, Gonzalez-Arroyo:2012ztz, Gonzalez-Arroyo:2013bta,Gonzalez-Arroyo:2013gpa, GarciaPerez:2013dgk, GarciaPerez:2015rda}. By using the minimal twist ($k=1$) and heavy adjoint fermions, one can obtain a one-site model that maintains $(\mathbb{Z}_N)^4$ center symmetry without the need for tuning $k/L$. The inclusion of a twist also allows us to define gauge-invariant observables with momentum- or position-space interpretation, leveraging the notion of emergent lattice coordinates in the reduced model. 

We first present the numerical evidence that, in the adjoint TEK model with $N=121$, the minimal twist $k=1$ and sufficiently heavy adjoint fermions ($\kappa=0.01$-$0.03$), the $(\mathbb{Z}_N)^4$ center symmetry remains unbroken over a broader region of the ($\kappa,b$) parameter space compared to the untwisted case.
The result also shows that the introduction of heavy adjoint fermions stabilizes the $(\mathbb{Z}_N)^4$ center-symmetric vacuum of the TEK model with small $k$ as $k/L < 1/9$, where the symmetry is broken in the absence of heavy adjoint fermions. 
Subsequently, we partially extend the one-site model along one lattice direction to realize a geometry akin to $\mathbb{R}^3 \times S^1$, allowing us to study the behavior of the theory under $S^1$ compactification with adjoint fermions having periodic boundary condition (periodic adjoint fermions). Our numerical simulations in this partially reduced TEK model for $N=16$-$49$, $b=0.30$-$0.46$ and $\kappa=0.03$-$0.16$ are consistent with the hypothesis of {\it adiabatic continuity} \cite{
Davies:1999uw, Davies:2000nw, Unsal:2007vu, Unsal:2007jx, Kovtun:2007py, Shifman:2008ja, Unsal:2010qh, Poppitz:2009uq, Anber:2011de, Poppitz:2012sw, Misumi:2014raa, Hongo:2018rpy,Misumi:2019dwq,Fujimori:2019skd, Misumi:2019upg, Fujimori:2020zka, Unsal:2020yeh, Poppitz:2021cxe, Tanizaki:2022ngt, Tanizaki:2022plm, Hayashi:2023wwi, Hayashi:2024qkm, Hayashi:2024gxv, Hayashi:2024yjc, Hayashi:2024yjc, Hayashi:2024psa, Guvendik:2024umd}: in the presence of periodic adjoint fermions, the theory remains in a center-symmetric (confined) phase as the $S^1$ circle is made small, smoothly connecting to the large-circle confining phase without a phase transition. 
This contrasts with the deconfinement transitions observed in pure partially reduced TEK models or those with antiperiodic adjoint fermions. 
We also in detail study the stability of the volume independence in the partially reduced model with the symmetric twist \cite{Gonzalez-Arroyo:2010omx} and compare the numerical results in the models with the symmetric twist and the modified twist \cite{Gocksch:1983iw,Gocksch:1983jj}.
Furthermore, we investigate the potential interplay between confinement and conformality in the small-mass region, where the theory may exhibit crossover behavior between confining and conformal regimes, influencing the interpretation of adiabatic continuity.

The organization of the paper is as follows. In Sec.~\ref{sec:TEK}, we review the twisted Eguchi-Kawai model and
introduce adjoint fermions to the model. In Sec.~\ref{sec:ZN}, we demonstrate by lattice Monte Carlo simulations that heavy adjoint fermions stabilize the $(\mathbb{Z}_N)^4$ center-symmetric vacuum in the TEK model with the minimal twist $k=1$. 
In Sec.~\ref{sec:FT} we introduce a partially reduced TEK model on $\mathbb{R}^3\times S^1$ with adjoint fermions with periodic boundary condition. 
In Sec.~\ref{sec:AC} we show that the numerical results provide support for the adiabatic continuity in the partially reduced TEK model on $\mathbb{R}^3\times S^1$ with periodic adjoint fermions, where the confined phase persists smoothly as the $S^1$ circle size is reduced. 
In Sec.~\ref{sec:SD}, we summarize our results and discuss future work.
In Appendix \ref{sec:SS}, we show the details of our simulations.
In Appendix \ref{sec:VI}, we show the numerical results on the volume independence in the partially reduced TEK models.

\section{Twisted Eguchi-Kawai model (with adjoint fermions)}
\label{sec:TEK}

\subsection{Twisted Eguchi-Kawai model}

The twisted Eguchi-Kawai (TEK) model we consider is defined by the gauge action
\begin{equation}
S_{\rm TEK} \,=\, -bN \sum_{\mu\neq\nu=1}^4 \mathrm{Tr}\left( z_{\mu\nu} U_\mu U_\nu U_\mu^\dagger U_\nu^\dagger \right)\,,
\label{eq:TEK}
\end{equation}
on a single-site lattice. Here $U_\mu$ are $SU(N)$ link matrices and 
\begin{align}
z_{\mu\nu} \,=\, \exp \left(\frac{2\pi i k}{L} \right)\,=\, z_{\nu\mu}^*\,,
\label{eq:twist}
\end{align}
are fixed twist phases. In the symmetric twist setup of Gonzalez-Arroyo and Okawa \cite{Gonzalez-Arroyo:2010omx}, one chooses $N=L^2$ and two integers $k,L$ such that $k$ and $L$ are relatively prime. In that case the twist is non-trivial in all $\mu\nu$ planes, and it can be shown that in the large-$N$ limit the $(\mathbb{Z}_N)^4$ center symmetry can remain intact even at sufficiently large $b$ (weak coupling), thus preserving the correspondence with the four-dimensional $SU(N)$ Yang-Mills theory \cite{Gonzalez-Arroyo:2010omx, Gonzalez-Arroyo:2012euf}. Essentially, the twisted background distributes the link eigenvalues in such a way as to minimize the free energy when $\langle\mathrm{Tr}\,U_\mu\rangle = 0$ for each $\mu$. However, maintaining this symmetry for finite but large $N$ requires careful selection of $k$ relative to $N=L^2$; one keeps the ratio $k/L$ fixed as $N$ increases. One of the proposals is to set $k/L$ to a ratio of every other Fibonacci numbers as $\displaystyle \sim \lim_{j\to \infty} F_{j} /F_{j+2} \to 0.382$ \cite{Chamizo:2016msz,Hayashi:2025doq}.

In our study, we implement a simpler twist setup: we take $k=1$ and $N$ be a perfect square as $N=L^2$, e.g. $N=121$. This means that the total twist flux is relatively small. By itself, such a small twist can not prevent $(\mathbb{Z}_N)^4$ center symmetry breaking for large $N$\cite{Gonzalez-Arroyo:2010omx, Gonzalez-Arroyo:2012euf, Azeyanagi:2007su}. But we will show that with heavy adjoint fermions, the $(\mathbb{Z}_N)^4$ center symmetry can remain stable even with this minimal twist $k=1$, as will be shown in Sec.~\ref{sec:ZN}. 
We emphasize that working with the minimal twist avoids having to tune $k$ as $N=L^2$ changes, which simplifies the model. 
Moreover, the presence of the non-zero twist still allows us to define momentum-space observables. For example, one can define a formal lattice of size $L$ embedded in the $U_\mu$ matrices, enabling the computation of correlators and masses in principle. In practice, we focus on bulk observables (Polyakov loops, etc.) in this work and leave hadronic observable measurements \cite{Gonzalez-Arroyo:2015bya} for future studies.

\subsection{Introduction of adjoint fermions}
\label{sec:AJ}

We include two flavors ($N_f=2$) of adjoint Wilson fermions with periodic boundary conditions in all directions. Since we will also consider a compactification later, we clarify here that we impose periodic boundary condition on adjoint fermions to preserve the center symmetry, as appropriate for zero-temperature or spatial-compactification setups. 
The fermion action on the one-site lattice is
\begin{equation}
S_F = \sum_{f=1}^{2}\bar\psi_f  D_W[U] \psi_f\,,
\label{eq:ajf}
\end{equation}
with $D_W$ being the Wilson-Dirac operator in the adjoint representation. It is written as
\begin{align}
D_W [U]= 1 - \kappa \sum_{\mu=1}^4 \big[(1-\gamma_\mu) U_\mu^{\rm adj} + (1+\gamma_\mu)U_\mu^{\rm adj,\dagger}\big]\,,     
\label{eq:do}
\end{align}
where $U_\mu^{\rm adj}$ are $N\times N$ matrices in the adjoint rep given by $(U_\mu^{\rm adj})_{ab} = 2\mathrm{Tr}(T_a U_\mu T_b U_\mu^\dagger)$ for $T_a$ the $SU(N)$ generators. The hopping parameter $\kappa$ is related to the bare adjoint quark mass $m_0$ as $m_0 = \frac{1}{2\kappa} - \frac{1}{2\kappa_c}$, where $\kappa_c$ is the critical value for which the quarks are massless (in practice $\kappa_c$ is determined by extrapolating the pion mass to zero). In this work we consider $\kappa$ well below $\kappa_c$, i.e. heavy (but dynamical) adjoint fermions that are not decoupled but also not near the massless limit. The presence of adjoint fermions provides a center-symmetry restoring force;
as discussed in the introduction, at strong coupling (small $b$) the naive Eguchi-Kawai action itself favors ${\mathbb Z}_N$-symmetric phase, while at weak coupling (large $b$) it prefers to ${\mathbb Z}_N$-broken phase. Heavy adjoint fermions effectively extend the ${\mathbb Z}_N$-symmetric regime towards weaker couplings by penalizing the ${\mathbb Z}_N$-broken configurations \cite{Bringoltz:2009kb, Bringoltz:2011by}. 

In this paper, we consider the $k=1$ twisted Eguchi-Kawai model with two-flavor adjoint fermions (adjoint TEK model)
\begin{align}
S_{\rm ATEK} = S_{\rm TEK}\,+\,\sum_{f=1}^{2} \bar\psi_f D_W[U]\psi_f\,, 
\end{align}
where $S_{\rm TEK}$ is the twisted Eguchi-Kawai action in Eq.~(\ref{eq:TEK}).

In the continuum, adjoint QCD with $N_f$ flavors has a one-loop $\beta$-function coefficient $b_0 = \frac{1}{24\pi^2}(11 - 4N_f)$ and a two-loop coefficient $b_1 = \frac{1}{192\pi^2}(17 - 16N_f)$. For $N_f=2$, $b_0$ remains positive (asymptotic freedom holds) but the two-loop coefficient $b_1$ becomes negative, suggesting the existence of an infrared fixed point if the fermions are massless. Indeed, $N_f=2$ adjoint QCD on the lattice has been shown to be in or near the boundary of the conformal window \cite{DelDebbio:2010zz,Athenodorou:2024rba}.
The conformality and the mass anomalous dimension in the adjoint TEK model have been intensively investigated in \cite{Gonzalez-Arroyo:2012ztz, Gonzalez-Arroyo:2013bta, Gonzalez-Arroyo:2013gpa, GarciaPerez:2013dgk, GarciaPerez:2015rda}. In our context of very heavy fermions, we are effectively dealing with a confining theory (closer to pure Yang-Mills in the infrared) in Sec.~\ref{sec:ZN}, while we will discuss possible influence of the conformality on the adiabatic continuity in the partially reduced model with relatively light adjoint quarks in Sec.~\ref{sec:AC}. We use the term “heavy” to mean the fermion mass is large enough that the system exhibits confinement and a discrete spectrum with the dynamical mass gap, not conformality. At the same time, the dynamical fermions are active enough to impact the $\mathbb{Z}_N$ center potential. This is the regime where we expect the adjoint TEK model to mimic large-$N$ pure gauge theory at long distances, as argued by Azeyanagi {\it et al.} \cite{Azeyanagi:2010ne}. 


\section{Numerical results of $\mathbb{Z}_N$ stability}
\label{sec:ZN}

\subsection{Polyakov loop and ${\mathbb Z}_N$-symmetric vacuum}

We mainly performed Monte Carlo simulations of $N_f=2$ adjoint TEK model for $N=121$ ($L=11$), $k=1$, $b=0.36$ and $\kappa=0.00$-$0.03  \ll \kappa_c \approx 0.17$. We use the Hybrid Monte Carlo (HMC) algorithm for the dynamical two-flavor adjoint Wilson fermions. The details of numerical calculations are shown in Appendix \ref{sec:SS}. To diagnose the $(\mathbb{Z}_N)^4$ center symmetry, we measure the absolute value of expectation value of Polyakov loop in each direction, 
\begin{align}
P_\mu = \frac{1}{N}|\langle \mathrm{Tr}U_\mu\rangle|\,.
\end{align}
In the $(\mathbb{Z}_N)^4$ symmetric phase one expects $P_\mu = 0$ (for all $\mu$), while in a $\mathbb{Z}_N$ broken phase, 
$\langle \mathrm{Tr}U_\mu\rangle$ tends to settle to one of the $N$-th roots of unity, giving $P_\mu>0$. 
Note that for $k/L = 1/11 < 1/9$ and $\kappa=0.00$ (TEK model without adjoint fermions), the $(\mathbb{Z}_N)^4$ center symmetry has been shown to be spontaneously broken \cite{Gonzalez-Arroyo:2010omx}, while for $k=0$ and $\kappa=0.00$-$0.03$ (adjoint EK model without twist), the symmetry is also spontaneously broken \cite{Bringoltz:2009kb, Bringoltz:2011by}.
Let us now examine the case where both $\kappa$ and $k$ are nonzero ($\kappa\not=0,\,k\not=0$).

\begin{figure}[t]
\includegraphics[width=8cm]{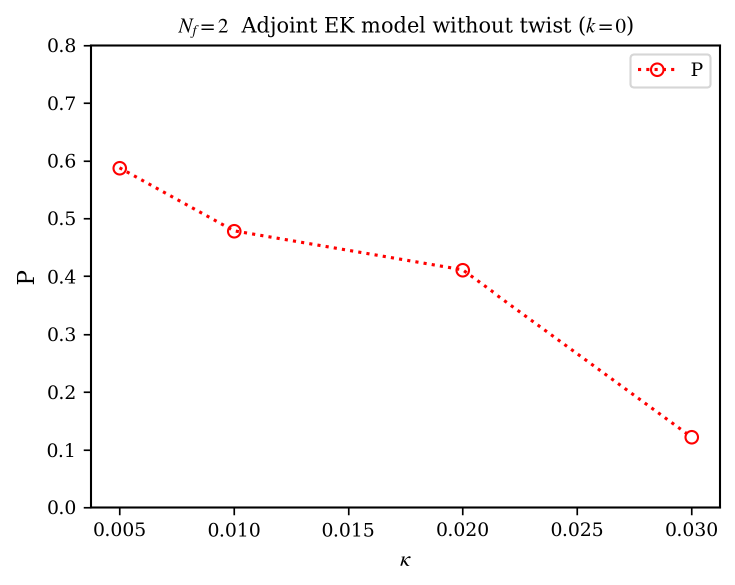}
\includegraphics[width=8cm]{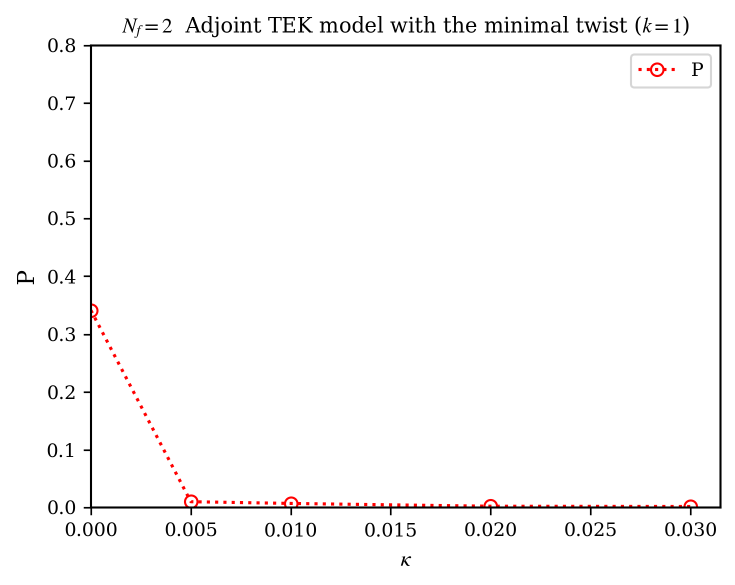}
\caption{Monte Carlo results for the magnitude of the Polyakov loop $P_\mu$ in the adjoint (T)EK model with $N=121$, $N_f=2$ and $b=0.36$ for $k=0,1$.
(Left) the results for $k=0$, namely the adjoint EK model without twist. (Right) the results for $k=1$, namely the adjoint TEK model with the minimal twist.}
\label{fig:P1}
\end{figure}

In Figure \ref{fig:P1}, we show Monte Carlo simulation results for the magnitude of the Polyakov loop $P_\mu$ in the adjoint (twisted) Eguchi-Kawai model with $N=121$ and $N_f=2$ at coupling $b=0.36$. The plot shows $P_\mu$ versus the hopping parameter $\kappa$ which controls the adjoint quark mass $\frac{1}{\kappa} - \frac{1}{\kappa_c}=m_q$, with the left figure for $k=0$ (untwisted) and the right figure for $k=1$ (minimally twisted). 

First of all, the results for the untwisted model ($k=0$) in the left figure exhibit $P_{\mu}\not=0$ in a small $\kappa$ region (large mass) and $P_\mu$ is getting smaller in a relatively large $\kappa$ region (relatively small mass), which is consistent with the results in the literature \cite{Bringoltz:2009kb, Bringoltz:2011by}. Note that the critical $\kappa$, which corresponds to zero adjoint quark mass, is $\kappa_{c}\sim 0.17$ for this parameter.

Surprisingly, the results for the minimally twisted model $(k=1)$ in the right figure show $P_{\mu}\sim 0$ from small $\kappa$ region (large mass) to large $\kappa$ region (relatively small mass), indicating an intact $({\mathbb Z}_N)^4$ center-symmetric phase, except $\kappa = 0.00$.
The case with $\kappa =0.00$ corresponds to the minimally twisted Eguchi-Kawai model without adjoint fermions, and our result for $\kappa =0.00$ is consistent with the known fact that the $({\mathbb Z}_N)^4$ center symmetry is broken for $k/N =1/11 <1/9$ \cite{Gonzalez-Arroyo:2010omx}. 
The key implication of this result is that the introduction of adjoint quarks into the TEK model with $k/N <1/9$ contributes to stabilizing $({\mathbb Z}_N)^4$ center symmetry even if the mass is very heavy.
In Appendix \ref{sec:SS}, we show the results for other $b$ including $b=0.35,0.40, 0.45$, which have qualitatively the same behavior as $b=0.36$.

\begin{figure}[t]
\includegraphics[width=8cm]{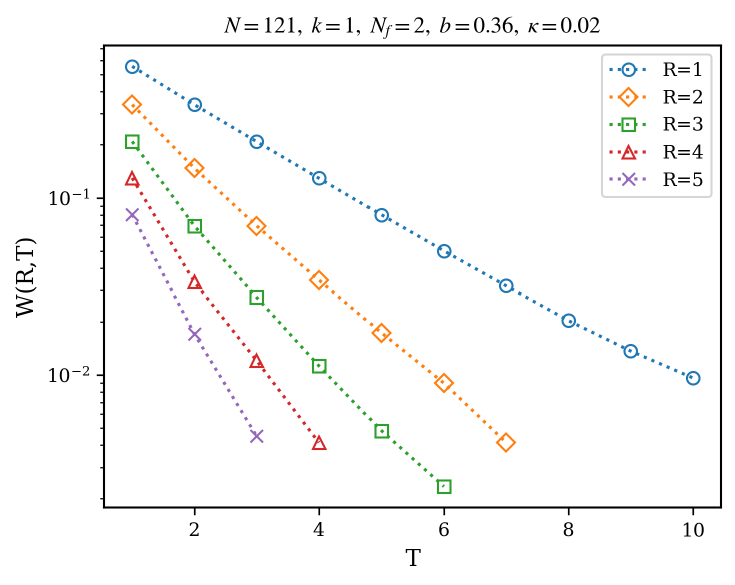}
\caption{Monte Carlo results of Wilson loop for different sizes of loops. The logarithm of the expectation values of the Wilson loops are roughly proportional to their areas, which suggest the area law and the existence of nonzero string tension.}
\label{fig:WL}
\end{figure}

In Figure \ref{fig:WL}, we show Monte Carlo results of the Wilson loop \cite{Gonzalez-Arroyo:2014dua} defined as
\begin{align}
W(R,T) \,=\, \langle{\rm Tr}\left(z_{\mu\nu}^{RT}(U_{\mu})^{R}(U_{\nu})^{T}(U_{\mu}^{\dagger})^{R}(U_\nu^{\dagger})^{T} \right)   \rangle \,,
\end{align}
with different loop sizes $(R,T)$ ($R,T\in {\mathbb N}$) in the adjoint TEK model with $N=121$, $N_f=2$, $b=0.36$ and $\kappa=0.02$. The plot shows that the logarithm of the expectation values of the Wilson loops are proportional to their areas, indicating the existence of the nonzero string tension and the area law. It means that the $({\mathbb Z}_N)^4$ symmetric phase we found in Fig.~\ref{fig:P1} is the confined phase, not the nearly conformal phase.

\subsection{Phase diagram in parameter space}

\begin{figure}[t]
\centering
\includegraphics[width=8cm]{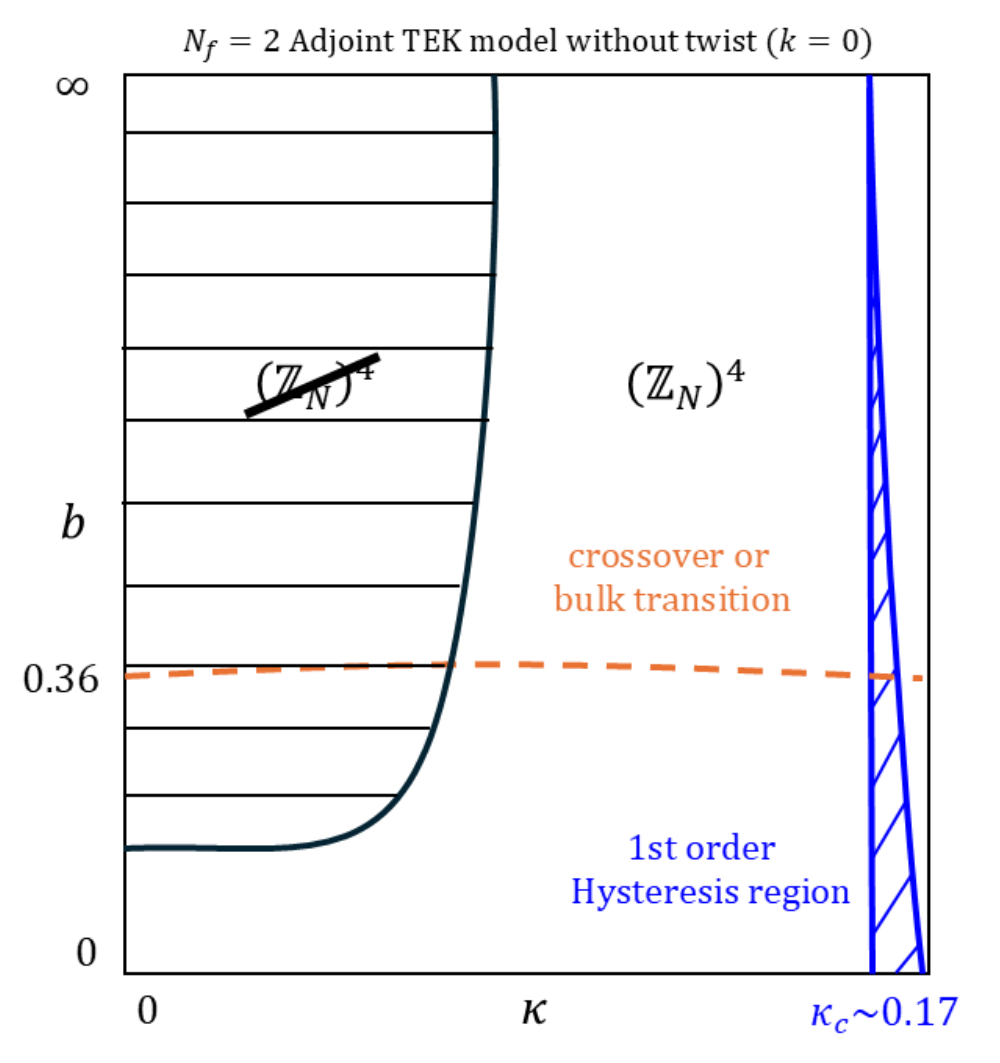}
\includegraphics[width=8cm]{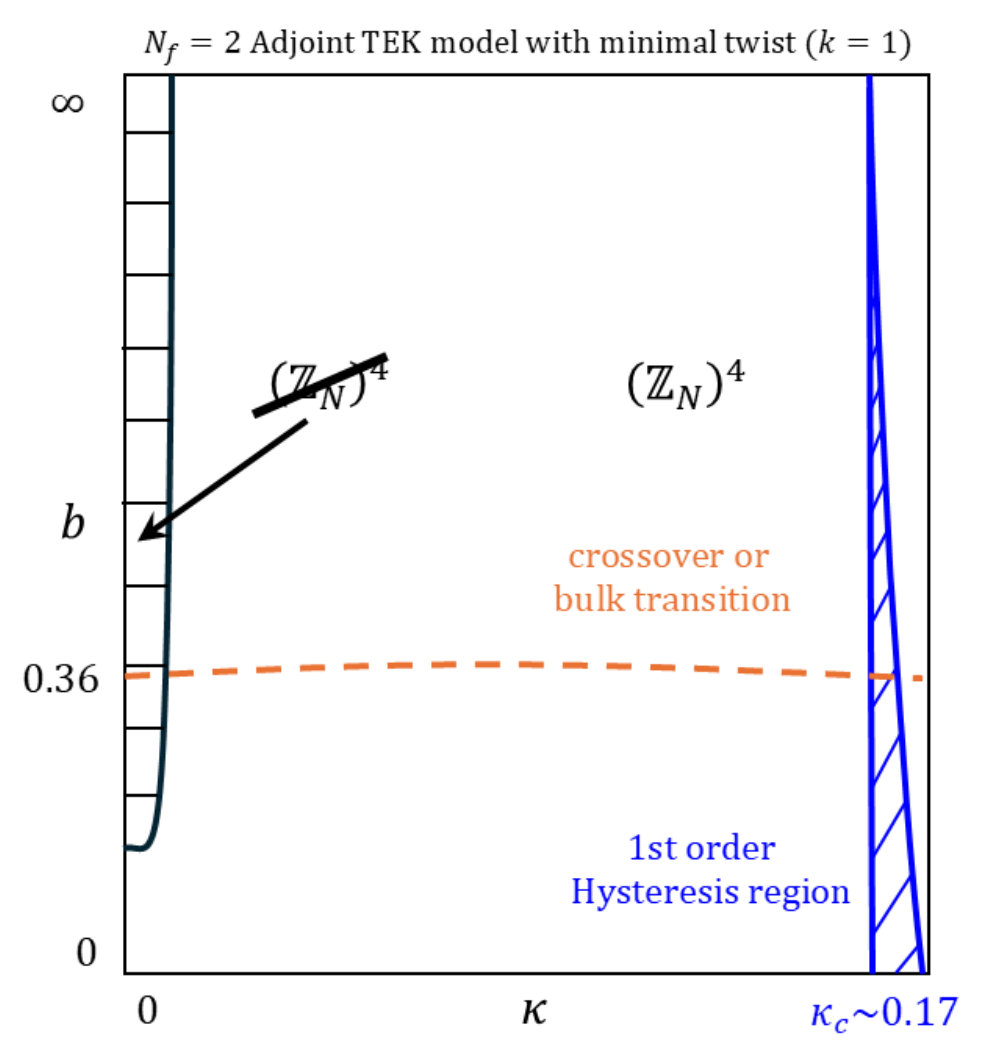}
\caption{Comparison of the schematic phase diagrams in the plane of gauge coupling $b$ and hopping parameter $\kappa$ for the adjoint EK model without twist (Left) and the adjoint TEK model with $k=1$ (Right).}
\label{fig:Phase1}
\end{figure}

In Figure \ref{fig:Phase1}, we show the schematic phase diagrams in the plane of gauge coupling $b$ vs. hopping parameter $\kappa$ for the adjoint EK model without twist in the left figure and the adjoint TEK model with $k=1$ in the right figure. 
In the left figure for the adjoint EK model without twist, if one focuses on the weak-coupling region (large $b$), $(\mathbb{Z}_N)^4$ center symmetry is unbroken only in the large $\kappa$ (small mass) region near $\kappa_c$ \cite{Bringoltz:2009kb, Bringoltz:2011by}. Note that the blue shaded region corresponds to a massless limit $\kappa_c$ including the 1st order hysteresis region. 
In the right figure for the adjoint TEK model with the minimal twist, we speculate that the $({\mathbb Z}_N)^4$ center symmetric phase extends to smaller $\kappa$ (large mass) region, although we need further investigation to show that the symmetric phase extends to $b=\infty$, or the continuum limit.
Our results in Fig.~\ref{fig:P1} at least indicate that the $({\mathbb Z}_N)^4$-broken region has shrunk significantly due to the introduction of the minimal twist into the adjoint EK model. 


What we are demonstrating is the trend that even heavy adjoint fermions push the system toward center symmetry. Our findings here corroborate and advance the earlier theoretical idea that adjoint fermions can successfully realize the large-$N$ volume independence without resorting to large twists \cite{Bringoltz:2009kb, Bringoltz:2011by, Azeyanagi:2010ne}. 
Having established the stability of the $(\mathbb{Z}_N)^4$ vacuum in the adjoint TEK model, we move on to an extension of this model to the one with one physical compactified dimension, to test the idea of continuity of the vacuum between small and large compactification radii.

\section{Partially reduced model on ${\mathbb R}^3 \times S^1$}
\label{sec:FT}

\subsection{Partially reduced TEK model}
\label{sec:PR}

One interesting application of volume independence is to finite-temperature or spatial compactification. 
Consider $SU(N)$ gauge theory on $\mathbb{R}^3 \times S^{1}$, where $S^1$ is a circle of circumference ${\mathcal L}_{4}$. If we take the $N\to\infty$ limit, a reduced model analogous to the TEK model can be formulated. In our context, we study a partially reduced TEK model with a lattice of size $1^3 \times L_4$, where three directions $\mu=1,2,3$ are one-site and one direction $\mu=4$ has $L_4$ sites. We apply ${\mathbb Z}_N$ twisted boundary conditions in the three reduced directions
and standard periodic boundary conditions in the extended direction. 
The action \cite{Gocksch:1983iw, Gocksch:1983jj, Azeyanagi:2010ne} is 
\begin{align}
S_{{\mathbb R}^3\times S^1}\, &=\, 
-bN\sum_{\tau=0}^{L_4 -1} \mathrm{Tr}\left(  \sum_{i=1}^{3}U_4(\tau) U_i(\tau+1) U_4^\dagger(\tau) U_i^\dagger(\tau)  + \sum_{i\neq j}^3  z_{ij} U_i(\tau) U_j(\tau) U_i^\dagger(\tau) U_j^\dagger(\tau) \right)+S_{F}\,,
\label{eq:R3S1}
\end{align}
where $\tau$ stands for lattice sites in the 4th direction with the size $L_4$ and $U_{4}(\tau)$ is the link variable along this direction while $U_{i}(\tau)$ ($i=1,2,3$) is the link variable along the three reduced directions with one site.
In this paper, we consider $L_4 \ll L=\sqrt{N}$.

It is notable that in the three directions $i=1,2,3$, $L^3 =N^{3/2}$ lattice effectively emerges due to the ${\mathbb Z}_N$ twist while the lattice size in the 4th direction is $L_4$, which is related to physical circumference as ${\mathcal L}_4 = L_4 a$, with $a$ being a effective lattice spacing.
Thus, the model can describe the lattice gauge theory on $L^3 \times L_4$ with $L_4 \ll L$ \cite{Azeyanagi:2010ne}.
This setup can be interpreted in two ways: (i) a finite-temperature system if anti-periodic boundary condition is imposed for the adjoint fermions in the 4th direction, or (ii) a spatial compactification if periodic boundary condition is imposed for the adjoint fermions in the 4th direction. We are interested in the case (ii) with periodic adjoint fermions, which is relevant to the concept of \emph{adiabatic continuity}. 

Adiabatic continuity, as conjectured by Unsal and others \cite{Unsal:2008ch}, suggests that $SU(N)$ gauge theory with adjoint matter or center-stabilized double-trace deformation on $\mathbb{R}^3\times S^1$ with the circumference ${\mathcal L}_{4}$ does not undergo a phase transition as ${\mathcal L}_{4}$ is varied, provided the fermions are periodic around $S^1$. In contrast, the same theory with thermal (anti-periodic) fermions have a deconfinement phase transition at some critical $T_c = 1/{\mathcal L}_{4}^{c}$. 

In the reduced model, the condition for volume independence on $\mathbb{R}^3 \times S^1$ is that the $(\mathbb{Z}_N)^3$ center symmetry in the three reduced directions remains unbroken. If that holds, the theory on the small $1^3\times L_4$ lattice is equivalent to the lattice gauge theory on $L^3 \times L_4$. 
The Wilson loop(line)
\begin{align}
W = \frac{1}{3N L_4} \sum_{\tau=0}^{L_4-1}\sum_{i=1}^3 \left|\langle\mathrm{Tr}U_i(\tau) \rangle\right|\,,
\label{eq:acW}
\end{align}
is an order parameter for the $(\mathbb{Z}_N)^3$ symmetry. 
Here $U_i(\tau)$ are the link matrices in the reduced directions, which can vary with the coordinate $\tau$ along the extended direction.
However, the condition of $W=0$ is necessary, but not a sufficient condition for the volume independence of the partially reduced model.
The necessary and sufficient condition for the volume independence is that all the possible Wilson lines defined as
\begin{align}
W(n_1, n_2, n_3) \equiv  \frac{1}{N L_4}  \sum_{\tau=0}^{L_4-1} \left|\langle\mathrm{Tr}[(U_1(\tau))^{n_{1}}(U_2(\tau))^{n_{2}}(U_3(\tau))^{n_{3}}] \rangle\right|\,,
\label{eq:wn123}
\end{align}
should be zero for arbitrary $-\sqrt{N} < n_1, n_2, n_3 < \sqrt{N}$.

Regarding our setup in Eq.(\ref{eq:R3S1}) with the symmetric twist in Eq.(\ref{eq:twist}), there has been an argument on the volume independence:
In Refs.\cite{Gocksch:1983iw, Gocksch:1983jj} it was discussed that the symmetric twist in the partially reduced TEK model without adjoint fermions could not lead to the correct large-$N$ gauge theory since the Wilson lines $W(n_1, n_2, n_3)$ can be nonzero and the volume independence breaks down in a large $b$ region. 
As we have discussed in the previous section, however, the introduction of adjoint fermions stabilizes the center symmetry and may strengthen the volume independence in the partially reduced TEK model.
Indeed, Ref.\cite{Azeyanagi:2010ne} shows that the model with the symmetric twist gives the deconfinement phase transition correctly.
In this paper, we consider two types of twists in the partially reduced models: one is the symmetric twist in Eq.(\ref{eq:twist}) and the other is the modified twist, which is given by
\begin{align}
&z_{12} \,=\, z_{23} \,=\, \exp \left(\frac{2\pi i l}{L} \right),\qquad
z_{13} \,=\, \exp \left(\frac{\pi i l}{L} \right),
\label{eq:mtwist}
\end{align}
with $l$ being an integer smaller than $L$, $z_{ji} = z_{ij}^{*}$ ($i=1,2,3$) and $N=L^2$. This type of the twist is originally proposed in Refs.\cite{Gocksch:1983iw, Gocksch:1983jj} to guarantee the volume independence as $W(n_1, n_2, n_3) = 0$.
In Sec.~\ref{sec:CS} and Sec.~\ref{sec:VI}, we exhibit the calculations to pursue the volume independence in the partially reduced TEK models with the symmetric twist and the modified twist. 
Then, in Sec.~\ref{sec:ac} and Sec.~\ref{sec:MT}, we investigate the adiabatic continuity in the setups with the symmetric twist and the modified twist. 

Meanwhile, the Polyakov loop around the $S^1$ (4th direction) is 
\begin{align}
P = \frac{1}{N}\left|\langle \mathrm{Tr}\prod_{\tau=0}^{L_4 -1} U_4(\tau)\rangle\right|\,.
\label{eq:acP}
\end{align}
Note that $P$ is not protected by any symmetry if the circle is spatial, so $P$ could in principle be non-zero even in $(\mathbb{Z}_N)^3$ symmetric phase. On the other hand, whether the system is in a physical confined phase or a deconfined phase is characterized by $P=0$ or $P\not= 0$ as long as the system is within the $(\mathbb{Z}_N)^3$ symmetric phase. In a thermal theory (with anti-periodic adjoint fermions), $P$ serves as the standard order parameter for deconfinement. Here, with periodic fermions it is an observable signaling adiabatic continuity vs. phase transition. In Sec.~\ref{sec:AC}, our simulations on the $1^3 \times L_4$ adjoint TEK model will use $N=16,25,36,49$ and $L_4=1$-$7$, $b=0.30$-$0.46$, $\kappa =0.03$-$0.16$. 
There, we will mainly measure $P$ as functions of $b$ for given $\kappa$ and $L_4$ to see if a deconfinement transition occurs. 
It is worthy noting that a larger (smaller) $b$ corresponds to a smaller (larger) circumference of compactification. 
We will also vary $L_4$ for given $\kappa$ and $b$ for the cross check in Sec.~\ref{sec:AC}.


Before going on to the next subsection, we make a comment on the parameter $\kappa=0.12$. In \cite{Gonzalez-Arroyo:2013gpa}, it was shown that the string tension in $N_f =2$ adjoint TEK model with $N=289$ and $k=5$ as a function of $\kappa=0.00$-$0.17$ starts to decrease at $\kappa\sim0.10$ and vanishes at the chiral limit $\kappa_c \sim 0.17$, indicating the conformality of the $N_f =2$ massless adjoint QCD.
This result suggests that $\kappa = 0.12$ may correspond to a marginal  mass in the confined-conformal crossover, although we use the different parameter set in this work.
In the next section, by giving a numerical reason, we argue that the adjoint TEK model with $\kappa=0.12$, or equivalently the partially reduced model with $\kappa=0.12$ and sufficiently small $b$ (corresponding to QCD(adj.) on ${\mathbb R}^4$) is in the confined phase and discuss the continuity of confined phase as the circumference is varied. Then, in the end of the section, we discuss the possibility of interplay between confined and nearly conformal regimes.


\subsection{Classical solution in the partially reduced model}
\label{sec:CS}
In this subsection, we analytically discuss how the volume independence can be broken in the partially reduced model in Eq.(\ref{eq:R3S1}) with the symmetric twist in Eq.(\ref{eq:twist}) by investigating the classical solutions \cite{Okawa:2025}.
In the limit $b \rightarrow \infty$, the link variable $U_{\mu}(\tau)$ that minimizes the action of the partially reduced TEK model with $1^3 \times L_4$ satisfies the following relation: 
\begin{align}
    U_i(\tau) U_j(\tau) &= \exp \left(\frac{2 \pi i k}{L} \right) U_j(\tau) U_i(\tau), \, \,\,(i < j) \label{con1} \\
    U_i(\tau + 1) &= U_4^{\dagger}(\tau) U_i(\tau) U_4(\tau)\,,
     \label{con2}
\end{align}
with $L^2 = N$ and $i,j=1,2,3$.
The link variable that satisfies this condition can be written as 
\begin{align}
    & U_1(\tau) = P_L \otimes D_1(\tau),  \\
    & U_2(\tau) = Q_L \otimes D_2(\tau),  \\
    & U_3(\tau) = Q_L P_L^{\dagger} \otimes D_3(\tau)\,,
\end{align}
where $P_L, Q_L$ are the $L \times L$ shift and clock matrices that satisfy the following relationship: 
\begin{equation}
    P_L Q_L =  \exp \left(\frac{2 \pi i k}{L} \right) Q_L P_L\,.
\end{equation}
Here, $D_i(\tau)$ are mutually commuting $L \times L$ matrices. 
From the condition in Eq.(\ref{con2}), $U_4(\tau)$ is written as 
\begin{equation}
    U_4(\tau) = {\bm 1}_L \otimes V(\tau)\,,
\end{equation}
where $V(\tau)$ is an arbitrary $L \times L$ matrix. 
Furthermore, if we define the gauge transformation function 
$\Omega(\tau) ={\bm 1}_{L}\otimes  V(1) V(2) \cdots V(\tau-1)$ with $\Omega(1) = \Omega(L_4 + 1) = {\bm 1}_N$, then we have 
\begin{equation}
\begin{split}
    U_4(\tau) \rightarrow \Omega(\tau) U_4(\tau) \Omega^{\dagger}(\tau + 1) &= \left({\bm 1}_L \otimes \prod_{s=1}^{\tau-1} V(s) \right) \left( {\bm 1}_L \otimes V(\tau) \right) \left({\bm 1}_L \otimes \prod_{s=1}^{\tau} V(s) \right)^{\dagger} \\ &= {\bm 1}_L \otimes {\bm 1}_L, 
    \label{U0_t}
\end{split}
\end{equation}
indicating that $U_{4}(\tau)$ can be made into an identity matrix ${\bm 1}_L \otimes {\bm 1}_L$ by the gauge transformation at least for $\tau=1,2,...,L_4 -1$. Regarding the final link variable $U_4(L_4)$, due to the periodic boundary conditions, it is written as
\begin{equation}
\begin{split}
    U_4(L_4)  & \rightarrow \Omega(L_4) U_4(L_4) \Omega^{\dagger}(L_4 + 1) \\ &= \left( {\bm 1}_L \otimes \prod_{s=1}^{L_4 - 1} V(s) \right) ( {\bm 1}_L \otimes V(L_4)) \\
    & = {\bm 1}_L \otimes \prod_{s=1}^{L_4} V(s)  = P_4 \,,
    \label{U0_L4}
\end{split}
\end{equation}
where $P_4$ is a Polyakov loop operator.
It means that $U_4 (L_4)$ can be identical to the Polyakov loop operator $P_4$, which can be a diagonal matrix. 

From these facts, we find that, even if ${\rm Tr}\,U_{\mu} = 0$, the trace of the special Wilson line ${\cal W}_{123}$ defined as 
\begin{equation}
    {\cal W}_{123} \equiv U_1(\tau) U_2(\tau)^{\dagger} U_3(\tau) = {\bm 1}_L \otimes (D_1(\tau) D_2^{\dagger}(\tau) D_3(\tau))\,,
\end{equation}
does not necessarily become zero. 
This is because, as shown in Eqs.(\ref{U0_t}) and (\ref{U0_L4}), $U_4(\tau)$ commutes with ${\cal W}_{123}$, so there is no restriction that forces ${\rm Tr}\,{\cal W}_{123} = 0$. On the other hand, in the model with the modified twist in Eq.~(\ref{eq:mtwist}), $U_1(\tau)$ and $U_3(\tau)$ do not commute with ${\cal W}_{123}$, which enforces the condition ${\rm Tr}\,{\cal W}_{123} = 0$ at least at the classical level
\footnote{In four dimensions $U_0(\tau)$ and ${\cal W}_{123}$ are non-commute, which enforces the condition ${\rm Tr}\,{\cal W}_{123} = 0$ even with the symmetric twist.}. 
We note $\frac{1}{NL_4}\sum_{\tau=0}^{L_4 -1}|\langle{\rm Tr}\,{\cal W}_{123}\rangle| = W(1,-1,1)$ in Eq.(\ref{eq:wn123}).

As seen from the discussion in this subsection, we need numerical study on not only $W$ in Eq.(\ref{eq:acW}), but also $W_{123}\equiv W(1,-1,1) =  \frac{1}{NL_4}\sum_{\tau=0}^{L_4 -1}|\langle{\rm Tr}\,{\cal W}_{123}\rangle|$ in the partially reduced model in Eq.(\ref{eq:R3S1}) to confirm the volume independence.
In the next subsection, we show the numerical results of $W_{123}$ to study the volume independence both in the models with the symmetric twist in Eq.(\ref{eq:twist}) and the modified twist Eq.(\ref{eq:mtwist}).


\subsection{Volume independence in partially reduced TEK models}
\label{sec:VI}

In the partially reduced TEK model in Eq.(\ref{eq:R3S1}) with the symmetric twist in Eq.(\ref{eq:twist}) without adjoint fermions, the volume independence can break down \cite{Gocksch:1983iw, Gocksch:1983jj}.
The incorporating adjoint fermions can, however, stabilize the $({\mathbb Z}_N)^3$ center symmetry on ${\mathbb R}^3$ and the volume independence may remain intact.
On the other hand, the model with the modified twist in Eq.(\ref{eq:mtwist}) is considered to be free from the breakdown of the volume independence \cite{Gocksch:1983iw, Gocksch:1983jj}.

As seen in the previous subsection, to investigate the problem in the partially reduced TEK models with the symmetric and the modified twists, we need to calculate the quantity $W_{123}$ besides $W$ \cite{Okawa:2025},
\begin{align}
W_{123} = \frac{1}{N L_4} \sum_{\tau=0}^{L_4-1} \left|\langle\mathrm{Tr}(U_1(\tau) U_2^{\dagger}(\tau) U_3(\tau) )\rangle\right|\,,
\end{align}
as the order parameter of the $(\mathbb{Z}_N)^3$ symmetry. 
In this subsection, we show the numerical Monte Carlo calculations for $\kappa=0.03,0.12$, $b=0.36$-$0.46$, $L_4 = 2$, $k=1$ ($l=1, 3$ for the modified twist in Eq.~(\ref{eq:mtwist})), $N_f=2$ and $N=16,36$.

Regarding the other order parameter of volume independence $W$ defined in Eq.(\ref{eq:acW}), we will show its numerical results along with $P$ when we investigate the adiabatic continuity in Sec.~\ref{sec:AC}.
So, we here focus on $W_{123}$ as the order parameter of the volume independence.

\begin{figure}[t]
\centering
\includegraphics[width=8cm]{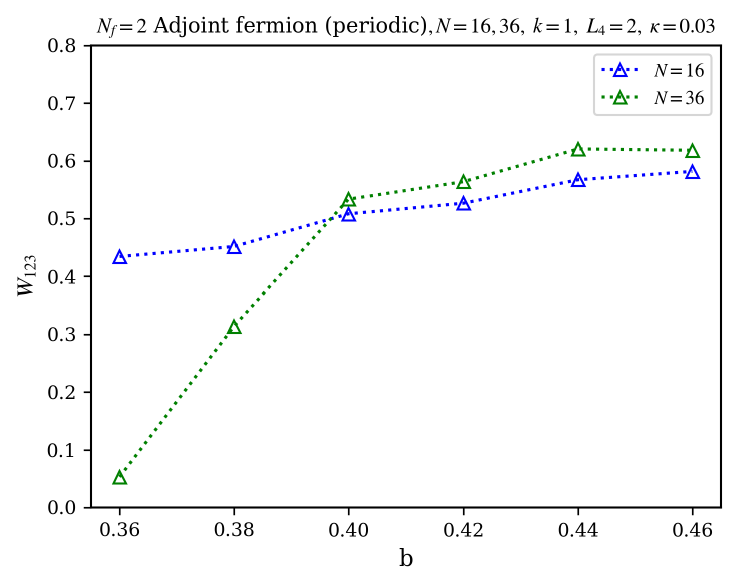}
\includegraphics[width=8cm]{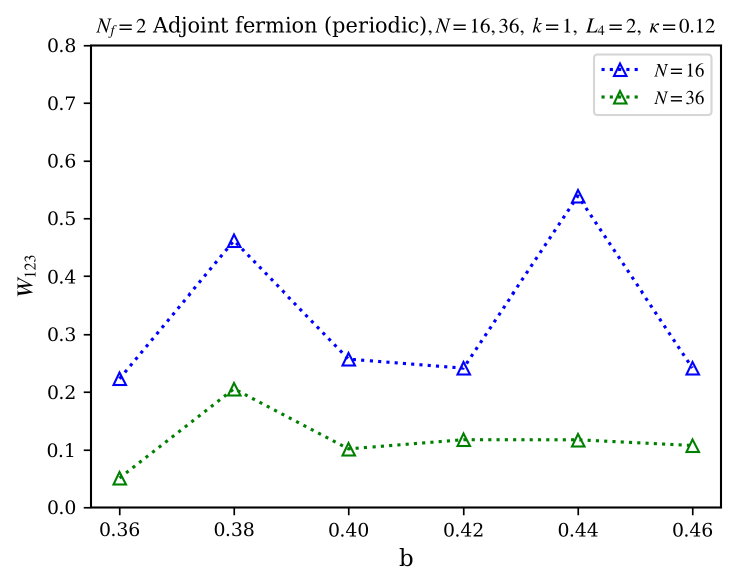}
\caption{Another volume-independence order parameter $W_{123}$ in the partially reduced TEK model with {\bf the symmetric twist}, $N=16,36$, $N_f=2$, $k=1$ and $L_4 = 2$ as a function of $b$. (Left) Heavy adjoint fermions $\kappa=0.03$: $W_{123}$ has nonzero values. (Right) Light adjoint fermions $\kappa=0.12$: $W_{123}$ gets smaller with larger $N$.}
\label{fig:VI}
\end{figure}

In Fig.~\ref{fig:VI}, we show the results of $W_{123}$ for the symmetric twist in Eq.(\ref{eq:twist}).
For heavy adjoint fermion with $\kappa=0.03$ (Left figure), $W_{123}$ has nonzero values and the $(\mathbb{Z}_N)^3$ symmetry seems to break down at least for heavy adjoint fermions.
For relatively light adjoint fermion with $\kappa=0.12$ (Right figure), the values of $W_{123}$ are fluctuating depending on $b$, but the values get smaller with increasing $N$. 
It suggests that the volume independence or the $({\mathbb Z}_N)^3$ center symmetry in the partially reduced TEK model with light adjoint fermions ($\kappa=0.12$) may recover in a large $N$. 
However, from these results with insufficient statistics and small $N$, we cannot give a conclusion on the volume independence with the symmetric twist. 
Thus, in the next section we will just assume the volume independence in the partially reduced model with the symmetric twist.

\begin{figure}[t]
\centering
\includegraphics[width=8cm]{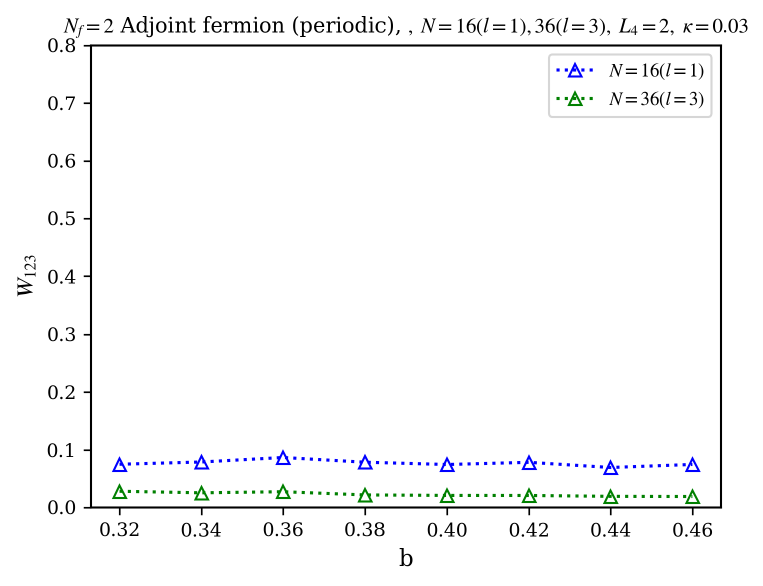}
\includegraphics[width=8cm]{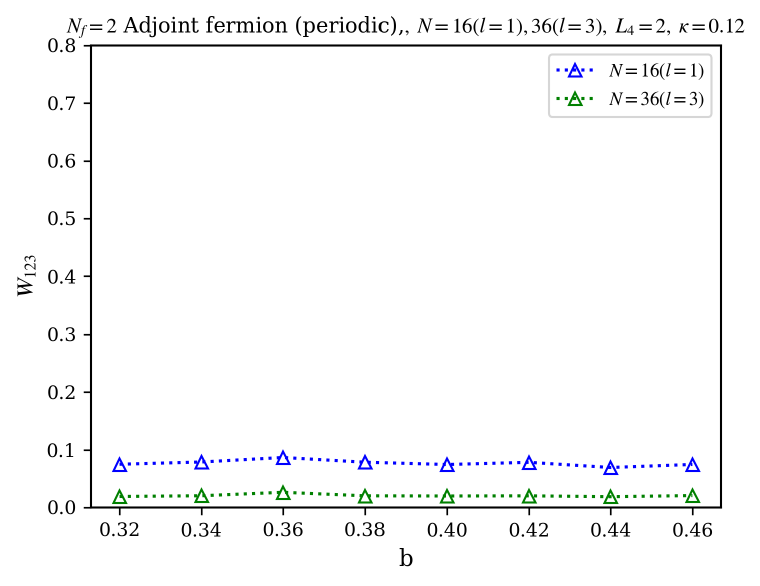}
\caption{Another volume-independence order parameter $W_{123}$ in the partially reduced TEK model with {\bf the modified twist}, $N=16\,(l=1),\,36\,(l=3)$, $N_f=2$,  and $L_4 = 2$ as a function of $b$. (Left) Heavy adjoint fermions $\kappa=0.03$:  $W_{123}$ remains near zero over the entire range. (Right) Light adjoint fermions $\kappa=0.12$: $W_{123}$ remains near zero over the entire range.}
\label{fig:VI2}
\end{figure}

In Fig.~\ref{fig:VI2}, we show the results of $W_{123}$ for the modified twist in Eq.(\ref{eq:twist}).
For both heavy ($\kappa=0.03$) and light ($\kappa=0.12$) adjoint fermions, $W_{123}$ remains near zero values and the $(\mathbb{Z}_N)^3$ symmetry seems to remain intact over the entire range. This tendency is stronger for larger $N$.
These results indicate that the volume independence exists in the partially reduced model with the modified twist.

\section{Numerical results of adiabatic continuity}
\label{sec:AC}

Before studying the adiabatic continuity in the partially reduced TEK model, let us remind ourselves of
the known facts on continuum gauge theory: for pure Yang-Mills (no fermions, or equivalently infinitely heavy adjoint fermions) or Yang-Mills theory with anti-periodic (thermal) adjoint quarks, one has a critical $T_c = 1/{\mathcal L}_{4}^{c}$ where $P$ jumps from zero to nonzero, indicating the standard confinement-deconfinement transition. 
By introducing adjoint fermions with periodic boundary conditions, this transition could be avoided.

\subsection{Deconfinement transition with anti-periodic adjoint fermions}

\begin{figure}[t]
\centering
\includegraphics[width=8cm]{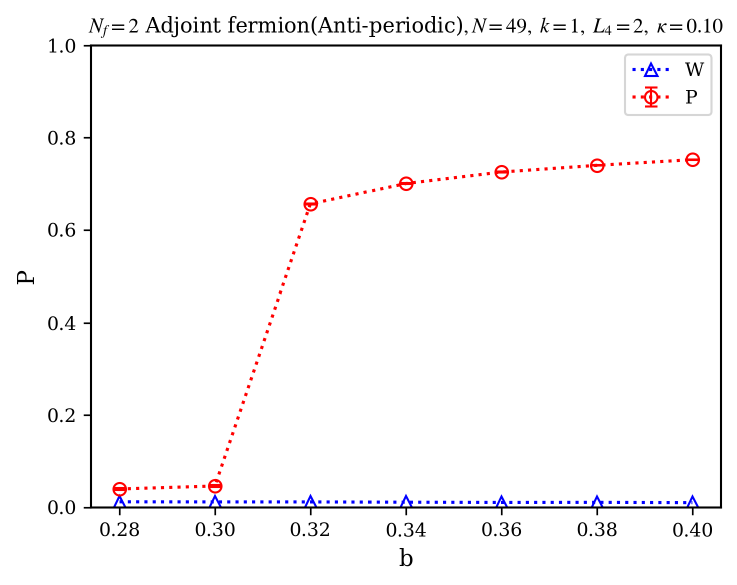}
\includegraphics[width=8cm]{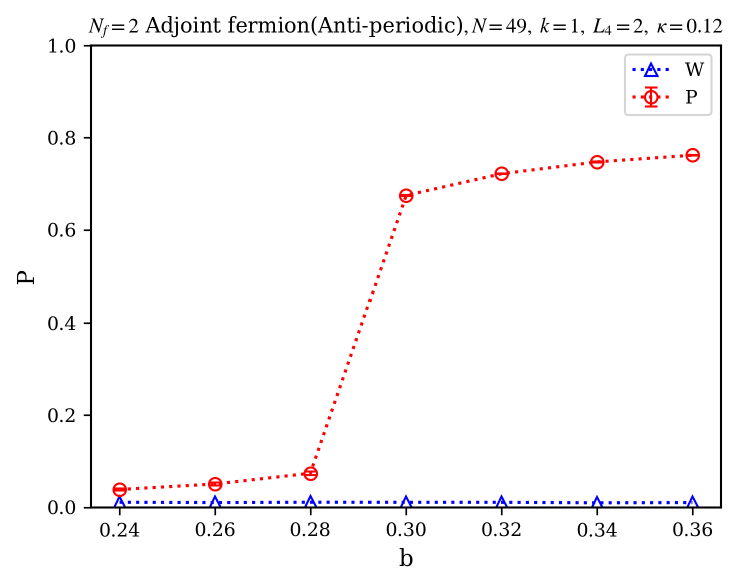}
\caption{Deconfinement order parameter $P$ (red circles) and the volume-independence order parameter $W$ (blue triangles) in the partially reduced TEK model with $N_f=2$ anti-periodic adjoint fermions, $N=49$, $k=1$ (symmetric twist), $L_4 = 2$, $\kappa=0.10$(Left) and $\kappa=0.12$(Right) as a function of $b$.}
\label{fig:APBC1}
\end{figure}

We first check the validity of our setup by studying the case with anti-periodic adjoint fermion (finite-temperature case), and comparing our results with the known result.
In Figure \ref{fig:APBC1}, the deconfinement order parameter $P$ in the partially reduced TEK model with two-flavor anti-periodic adjoint fermions, $N=49$, $k=1$(symmetric twist) and $L_4 = 2$ for $\kappa=0.10$ (left figure)  and $\kappa=0.12$ (right figure) are depicted as a function of $b$. 
While the spatial Polyakov loop $W$ remains zero, the $S^1$ Polyakov loop $P$ shows a sharp rise around $b=0.30$ (left figure) for $\kappa=0.10$ and $b=0.28$ (right figure) for $\kappa=0.12$, indicating deconfinement phase transition, consistent with the result in Ref.~\cite{Azeyanagi:2010ne}. 
This results also suggest that the model with the parameter set $N=49$, $k=1$, $\kappa=0.12$ and sufficiently large circumference (small $b$) is in the confined phase, not in the nearly conformal phase.

\begin{figure}[t]
\centering
\includegraphics[width=8cm]{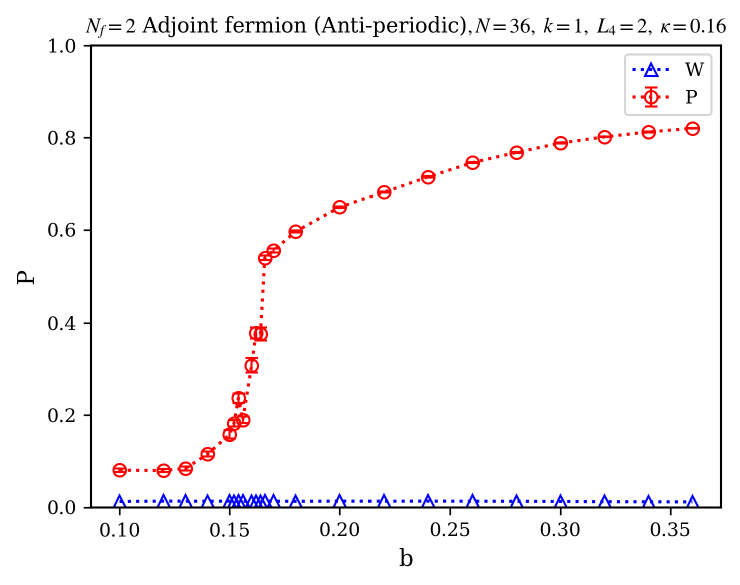}
\caption{Deconfinement order parameter $P$ (red circles) and the volume-independence order parameter $W$ (blue triangles) in the partially reduced TEK model with $N_f=2$ anti-periodic adjoint fermions, $N=36$, $k=1$ (symmetric twist), $\kappa=0.16$ and $L_4 = 2$ as a function of $b$.}
\label{fig:APBC2}
\end{figure}

In Figure \ref{fig:APBC2}, we also show $P$ as a function of $b$ in the partially reduced TEK model with two-flavor anti-periodic adjoint fermions, $N=36$, $k=1$ and $L_4 = 2$ for $\kappa=0.16$, which is very close to the massless (chiral) limit $\kappa=\kappa_c = 0.17$.
This result also indicates the phase or crossover transition from confinement to deconfinement phase.
The detailed phase diagram for QCD(adj.) with anti-periodic adjoint fermions is discussed in the literature (see \cite{Misumi:2014raa} for example).
We now move on to the main results: {\it adiabatic continuity}.


\subsection{Adiabatic continuity with the symmetric twist}
\label{sec:ac}

We now observe that for relatively light adjoint quarks ($\kappa=0.12 < \kappa_c = 0.17$) with periodic boundary condition and the symmetric twist in Eq.(\ref{eq:twist}), 
no abrupt changes in $P$ are seen as $b$ (effective circumference) is varied. 

\begin{figure}[t]
\centering
\includegraphics[width=8cm]{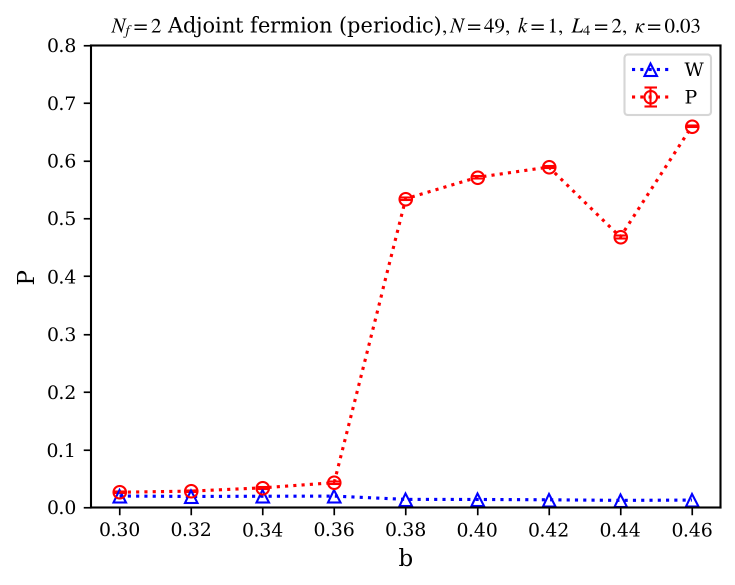}
\includegraphics[width=8cm]{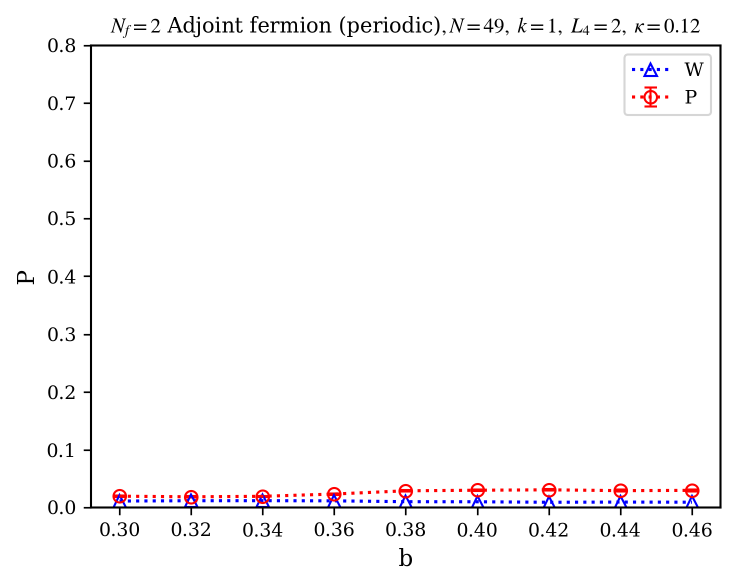}
\caption{Deconfinement order parameter $P$ (red circles) and the volume-independence order parameter $W$ (blue triangles) in the partially reduced TEK model with the symmetric twist, $N=49$, $N_f=2$ (periodic b.c.), $k=1$ and $L_4 = 2$ as a function of $b$. (Left) Heavy adjoint fermions $\kappa=0.03$: $W$ remains zero while $P$ shows a sharp rise around $b=0.36$, indicating deconfining phase transition. (Right) Lighter adjoint fermions $\kappa=0.12$: both $W$ and $P$ remain near zero over the entire range, with no sign of a phase transition. }
\label{fig:AC1}
\end{figure}

\begin{figure}[t]
\centering
\includegraphics[width=8cm]{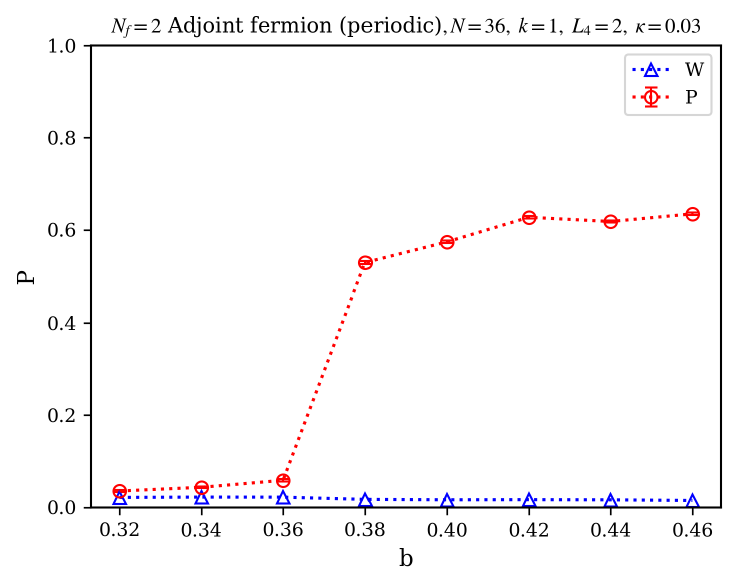}
\includegraphics[width=8cm]{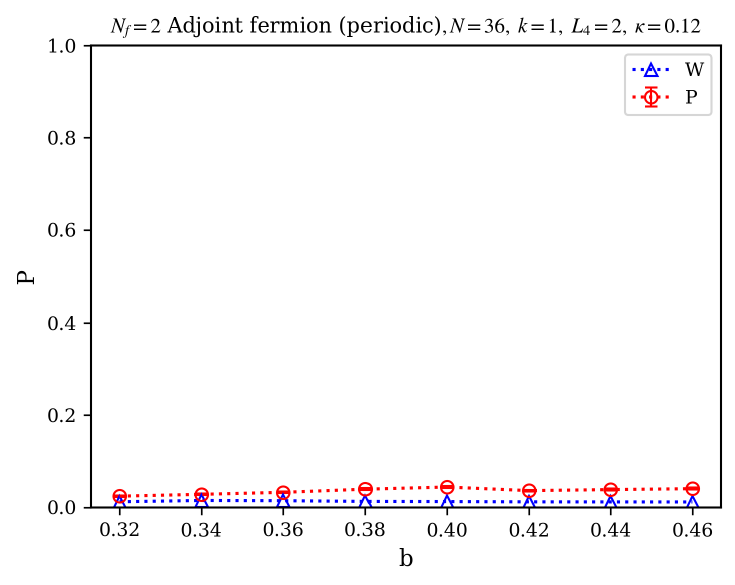}
\caption{Deconfinement order parameter $P$ (red circles) and the volume-independence order parameter $W$ (blue triangles) in the partially reduced TEK model with the symmetric twist, $N=36$, $N_f=2$, $k=1$ and $L_4 = 2$ as a function of $b$. (Left) Heavy adjoint fermions $\kappa=0.03$:  $W$ remains zero while $P$ shows a sharp rise around $b=0.36$. (Right) Lighter adjoint fermions $\kappa=0.12$: $W$ and $P$ remain zero over the entire range. }
\label{fig:AC2}
\end{figure}

In Fig.~\ref{fig:AC1}(Left), with $\kappa=0.03$ (heavy quark), the quantity $P$ clearly departs from zero at $b\approx0.36$, signaling the phase transition for the small $S^1$, while $W$ remains zero for all values of $b$. 
This demonstrates that if adjoint fermions are too heavy, close to pure gauge theory, the model shows a deconfinement transition at certain critical coupling $b_c$, effectively corresponding to the critical circumference of compactification. 
Note that the value of Polyakov loop $P$ at $b=0.44$ gets smaller  compared to other values of $b$ in Fig.~\ref{fig:AC1}(Left).
We consider that this behavior likely originates from the fact that the spontaneous breaking of ${\mathbb Z}_{N}$ center symmetry has various patterns such as ${\mathbb Z}_{N}\to{\mathbb Z}_{\ell}$ ($\ell<N$).

In contrast, Fig.~\ref{fig:AC1}(Right) with $\kappa=0.12$ (lighter quark) shows $W\approx0$ for all $b$ in the range, and  no noticeable jump in $P$; rather, $P$ stays very small. This is consistent with the absence of any transition -- i.e. suggesting that the theory remains in the confined phase even when the spatial compactification circumference is quite small. Here, $L_4=2$ with large $b$ (effectively small lattice spacing) corresponds to a rather small compactification circumference. 
We also show the similar results for $N=36$ in Figure \ref{fig:AC2}, where we obtain qualitatively the same behavior of $P$ (no phase transition) as $N=49$. 

\begin{figure}[t]
\centering
\includegraphics[width=8cm]{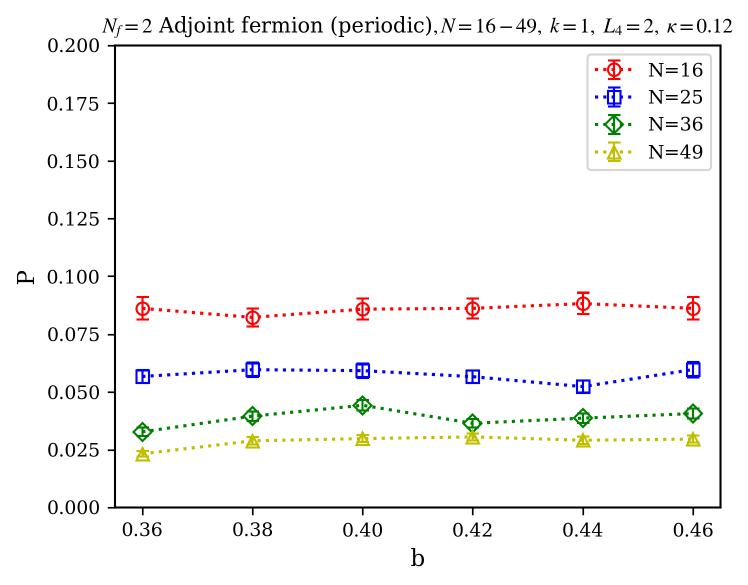}
\caption{$N$-dependence of the Polyakov loop $P$ in the partially reduced TEK model with $N_f=2$ periodic adjoint fermions, $k=1$ (symmetric twist) and $L_4 = 2$ for $\kappa=0.12$. Larger $N$ means larger lattice size $L^3 L_4$. The nonzero value $P$ approaches to zero for larger $N$.}
\label{fig:FVE}
\end{figure}

To investigate the finite-volume effect on the Polyakov loop $P$, we show the dependence of $P$ on $N$ in the partially reduced TEK model with fixing $N_f=2$ (periodic b.c.), $k=1$ and $L_4 = 2$ for $\kappa=0.12$ in Fig.~\ref{fig:FVE}.
Larger $N=L^2$ means larger lattice size $L^3 L_4$. One finds that the small but nonzero values of $P$ approaches zero for larger $N=L^2$, indicating that these nonzero values of $P$ originate from finite-volume (finite-$N$) effects and are expected to vanish for large volume or large $N$ limits. It reinforce the statement that the right figures in Fig.~\ref{fig:AC1} and Fig.~\ref{fig:AC2} exhibit the absence of deconfinement transition.

There is another possibility: a small but nonzero value of $P$ may indicate the partial breaking of the ${\mathbb Z}_N$ symmetry to ${\mathbb Z}_{N-1}$ subgroup. For such a case, we have $P\approx {\mathcal O}(1/N)$, and in the large $N$ limit $P$ goes to zero, which is also consistent with the behavior of $P$ in Fig.~\ref{fig:FVE}.
In this scenario, precisely speaking, the adiabatic continuity is broken down, but one cannot distinguish ${\mathbb Z}_N$ and ${\mathbb Z}_{N-1}$ phases in the large $N$ limit.
In this sense, the adiabatic continuity {\it effectively} exists in the large $N$ limit even for this scenario.

\begin{figure}[t]
\centering
\includegraphics[width=8cm]{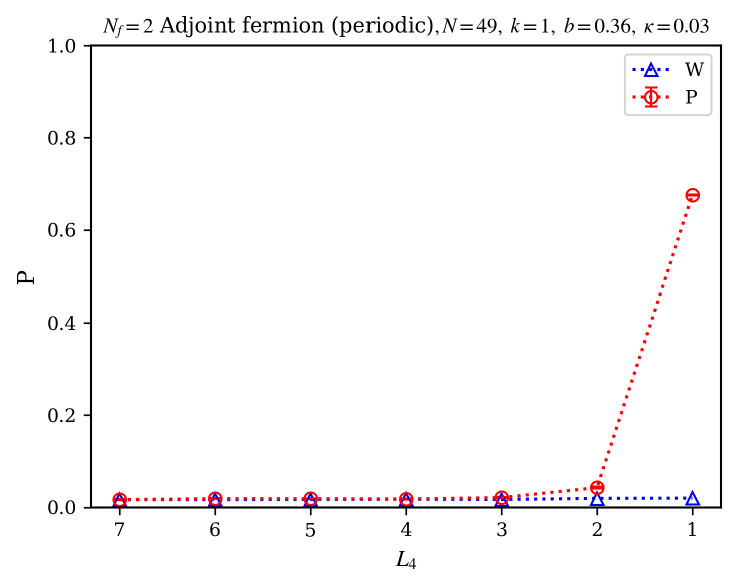}
\includegraphics[width=8cm]{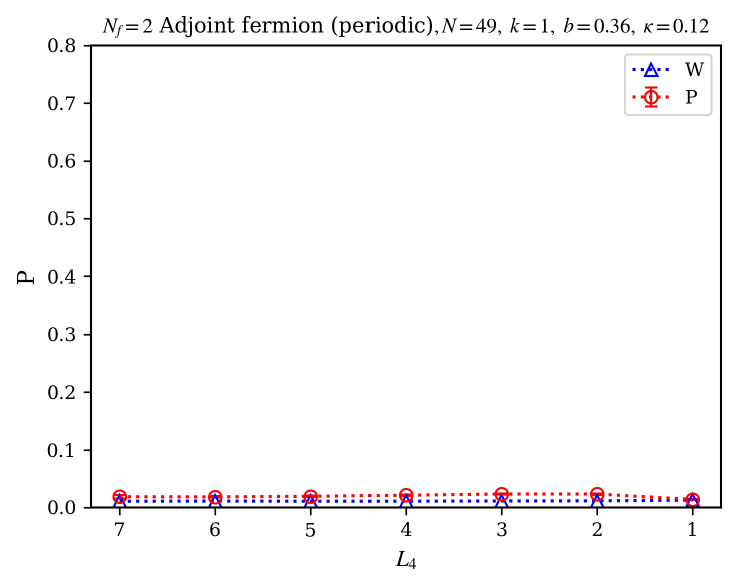}
\caption{$P$ (red circles) and $W$ (blue triangles) in the partially reduced TEK model with the symmetric twist ($k=1$), $N=49, b = 0.36$ and $N_f=2$ as a function of $L_4$. (Left) Heavy adjoint fermions $\kappa=0.03$: $W$ remains zero while $P$ shows a sharp rise around $L_4= 2$. (Right) Lighter adjoint fermions $\kappa=0.12$: both $W$ and $P$ remain near zero over the entire range, with no sign of a phase transition. }
\label{fig:PW}
\end{figure}

As further investigation, we vary $L_4$ from $1$ to $7$ with $N=49$ and $b=0.36$ fixed, and calculate the Polyakov loop $P$ for a heavy quark case ($\kappa=0.03$) and a lighter quark case ($\kappa=0.12$) in Figure \ref{fig:PW}. In Fig.~\ref{fig:PW}(Left) with $\kappa=0.03$ (heavy quark), while the quantity $W$ is consistent with zero for $L_4 = 1$-$7$, $P$ rises from near zero to a large value at $L_4 = 2$, indicating a deconfinement transition for the small $S^1$. This demonstrates that if adjoint fermions are too heavy, the model shows a finite-temperature transition at some compactification circumference. In contrast, Fig.~\ref{fig:PW}(Right) with $\kappa=0.12$ (lighter quark) shows $W\approx0$ for all $L_4$ in the range, and $P$ also stays very small up to the smallest $L_4$. 
These are also consistent with the absence of deconfinement transition.

To sum up, our results on the partially reduced model with the symmetric twist are consistent with the adiabatic continuity conjecture: the small-circle confining theory with periodic adjoint fermions continuously connects to the large-circle confining phase.
However, we also have to note that the volume independence may be broken in the model with the symmetric twist as shown in Sec.~\ref{sec:VI}.


In Section \ref{sec:VI}, the detailed calculations on the volume independence in the partially reduced TEK model with the symmetric twist are exhibited in Fig.\ref{fig:VI}, where we find the indication of the volume independence for $\kappa=0.12$ (light fermions) in a large $N$, although we cannot draw a conclusion on it. 
On the other hand, we confirm it in the model with the modified twist, inspired by Ref.\cite{Gocksch:1983iw, Gocksch:1983jj}, in Fig.\ref{fig:VI2}.
So, in the next subsection, we show the numerical results in the partially reduced TEK model with the modified twist, strongly suggesting the adiabatic continuity too.

\subsection{Adiabatic continuity with the modified twist}
\label{sec:MT}

In this subsection, we adopt the modified twist in Eq.(\ref{eq:mtwist}) in the partially reduced TEK model with periodic adjoint fermions \footnote{The twist in Eq.~(\ref{eq:mtwist}) is a further modified one from originally proposed one in Refs.\cite{Gocksch:1983iw, Gocksch:1983jj}}.
We here adopt $N = 36$, and set $l=3$ in the modified twist in Eq.(\ref{eq:mtwist}). It is worthy noting that, for large $N$, $l$ must be increased to maintain $(\mathbb{Z}_N)^3$ center symmetry. 
In Fig.\ref{fig:VI2} of Sec~\ref{sec:VI}, we show the numerical evidence of the volume independence in the partially reduced model with this modified twist.

\begin{figure}[t]
\centering
\includegraphics[width=8cm]{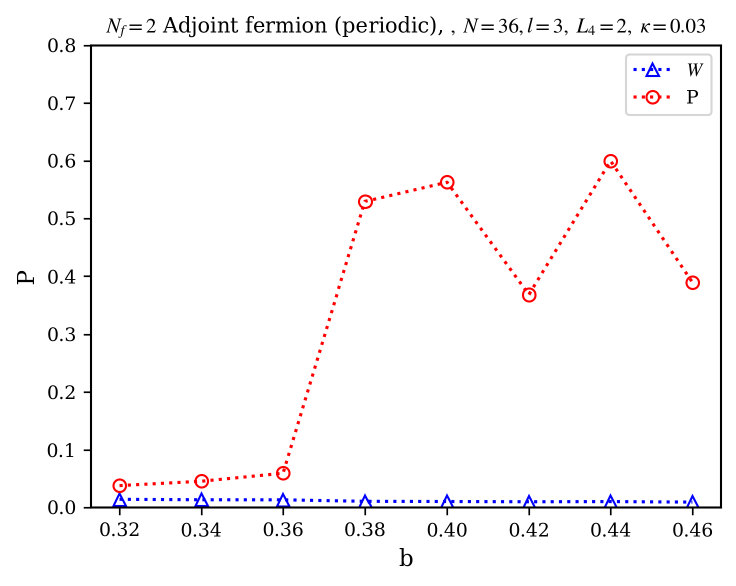}
\includegraphics[width=8cm]{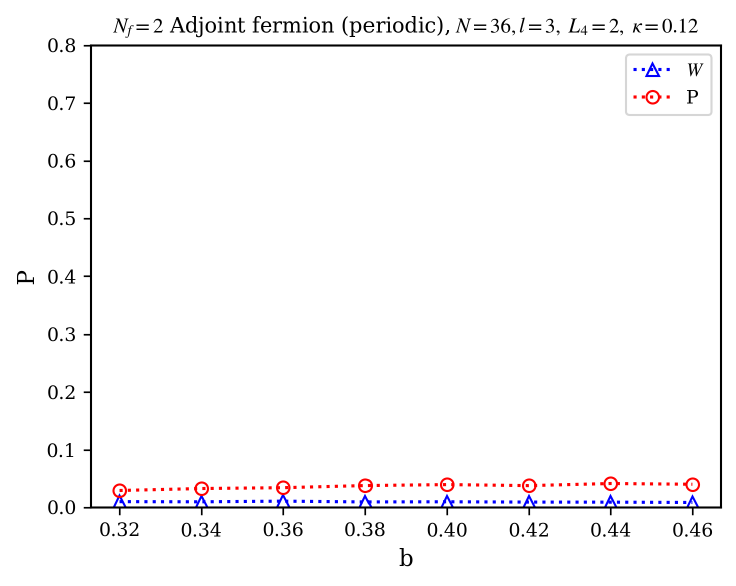}
\caption{$P$ (red circles) and $W$ (blue triangles) in the partially reduced TEK model with the modified twist, $N=36$, $N_f=2$, $l=3$ and $L_4 = 2$ as a function of $b$. (Left) Heavy adjoint fermions $\kappa=0.03$:  $W$ remains zero while $P$ shows a sharp rise. (Right) Lighter adjoint fermions $\kappa=0.12$: both $W$ and $P$ remain near zero. }
\label{fig:MT}
\end{figure}

In Fig.~\ref{fig:MT}, we show the numerical results of the deconfinement order parameter $P$ and the volume independence order parameter $W$ with $\kappa=0.03$ (heavy quark) and $\kappa=0.12$ (light quark) in the partially reduced TEK model with the modified twist in Eq.(\ref{eq:mtwist}), $N=36$, $N_f=2$ (periodic b.c.) and $L_4 = 2$ as a function of $b$.

In the left figure of Fig.~\ref{fig:MT} for $\kappa=0.03$, $P$ clearly departs from zero, signaling the phase transition for the small $S^1$, while $W$ remains zero. 
In the right figure of Fig.~\ref{fig:MT} for $\kappa=0.12$,
there is no jump in $P$ staying small. It means the absence of transition, suggesting that the theory remains in the confined phase when the compactification circumference is varied from large to small. As we have discussed in Fig.~\ref{fig:FVE}, small but nonzero values of $P$ for $\kappa=0.12$ in the right figure of Fig.~\ref{fig:MT} originate from the small-$N$ effects and will disappear in a large $N$. We obtain the strong support of adiabatic continuity even in the model with the volume independence being confirmed.


\subsection{Phase diagram: Confining vs Conformal}
\label{sec:PDCC}

So far, we have considered that the adjoint TEK model with $\kappa=0.12$, or equivalently the partially reduced model with $\kappa=0.12$ and sufficiently small $b$ (large circle) is in the confined phase and argued that Figures \ref{fig:AC1}, \ref{fig:AC2}, \ref{fig:PW} and \ref{fig:MT} indicate the adiabatic continuity of confined phases.
It is however worth noting that with $\kappa=0.12$, the system may cross into a regime where the theory is regarded as nearly conformal field theory deformed by nonzero fermion mass \cite{Gonzalez-Arroyo:2013gpa}. The numerical results of string tension and mass anomalous dimension for $N_f=2$ adjoint TEK model have suggested that $N_f=2$ adjoint QCD is conformal in the chiral limit, consistent with the lattice QCD simulations \cite{DelDebbio:2010zz}. Although our simulations with $\kappa=0.12$ does not seem to reach a point where conformality is manifest and we consider that the theory remains confining, we cannot exclude the possibility that the theory is in nearly conformal regime, to which the confining regime is connected by a smooth crossover \cite{Poppitz:2009uq}. 

\begin{figure}[t]
\centering
\includegraphics[width=11cm]{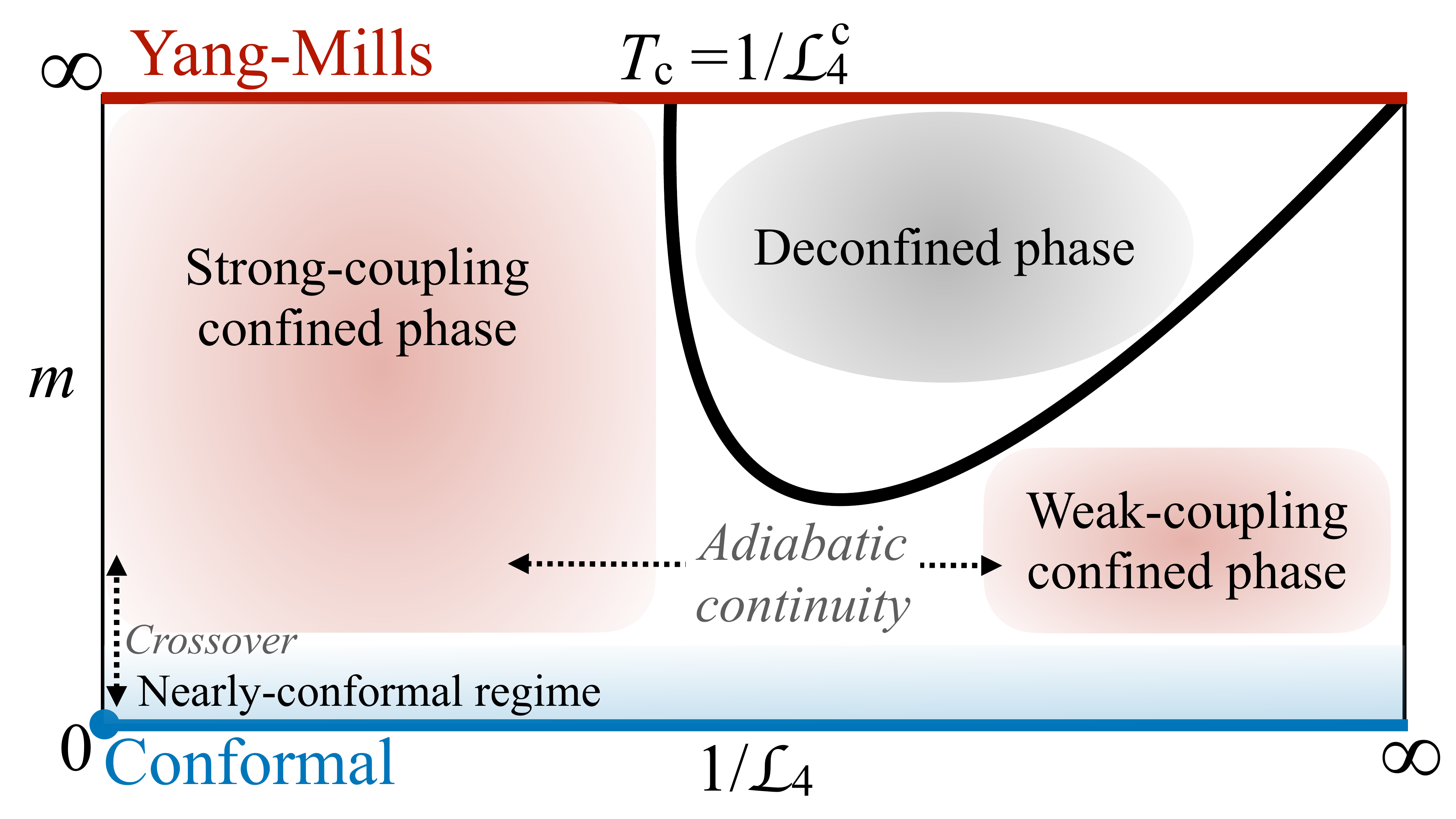}
\caption{Schematic illustration of ``confining" adiabatic continuity scenario for $SU(N)$ gauge theory with $N_f=2$ periodic adjoint fermions on ${\mathbb R}^3 \times S^1 $ with the circumference ${\mathcal L}_{4}$. }
\label{fig:ACP}
\end{figure}

In Figure \ref{fig:ACP}, we show the schematic illustration of the adiabatic continuity scenario for $SU(N)$ gauge theory with $N_f=2$ periodic adjoint fermions on ${\mathbb R}^3 \times S^{1}$ with the circumference ${\mathcal L}_{4}$. In the scenario, in pure Yang-Mills ($m\to \infty$), as $1/{\mathcal L}_{4}$ increases, one encounters a critical point $T_c = 1 / {\mathcal L}_{4}^{c}$
where the Polyakov loop in the $S^1$ direction becomes non-zero. For lighter periodic adjoint fermions (intermediate $m$), to which we consider $\kappa=0.12$ corresponds, the system remains confined for all $1/{\mathcal L}_{4}$, with no phase transition separating large ${\mathcal L}_{4}$ confined phase (strong-coupling confine phase) from small ${\mathcal L}_{4}$ confined phase (weak-coupling confined phase) generated by the bion confinement mechanism \cite{Unsal:2007vu, Unsal:2007jx,Kovtun:2007py, Shifman:2008ja, Misumi:2014raa, Poppitz:2021cxe, Hayashi:2024psa}. For very light adjoint quarks ($m\sim 0$), the system could behave as nearly conformal field theory, namely mass-deformed and $S^1$-deformed conformal field theory, to which the strong-coupling confined phase is connected by a smooth crossover.
Our numerical results in Figures \ref{fig:AC1}, \ref{fig:AC2}, \ref{fig:PW} and \ref{fig:MT} strongly support this phase diagram.

We here comment on the issue of {\it confined} versus {\it conformal} in QCD(adj.) with $N_{f}=2$ on ${\mathbb R}^3\times S^1$. 
A semiclassical analysis suggests that $N_{f}=2$ QCD(adj.) on ${\mathbb R}^3\times S^1$ is in a confined phase at asymptotically large distances due to the Coulomb gas of bions \cite{Poppitz:2009uq}, even if the theory on ${\mathbb R}^4$ is within the conformal window. 
However, as a remnant of the conformality, the chiral symmetry breaking is absent in this ``special" confined phase, and furthermore it is quite difficult to observe this confinement in the realistic lattice simulations since it only emerges at sufficiently large distances \cite{Poppitz:2009uq}.
So, we cannot abandon the possibility that the system may be in a nearly conformal phase or in such a special confined phase with the remnant of conformality even for intermediate $m$ (e.g. $\kappa=0.12$), and the continuity we observed could be the adiabatic continuity of these phases or the crossover between them.
The model with $N_f =1$ has no such problem and the confined phase extends to the massless limit. The investigation of $N_f=1$ case will be performed in the future work.


\section{Summary and Discussion}
\label{sec:SD}

In this paper, we presented a study of the twisted Eguchi-Kawai (TEK) model with two-flavor adjoint fermions, focusing on two main aspects: (1) the stabilization of the $(\mathbb{Z}_N)^4$ center symmetry by heavy periodic adjoint quarks in the one-site TEK model with the minimal twist ($k=1$), and (2) the realization of adiabatic continuity in a partially reduced model on $\mathbb{R}^3\times S^1$ with periodic adjoint quarks. 

Our Monte Carlo results demonstrate that incorporating two flavors of heavy adjoint Wilson fermions indeed stabilizes the $(\mathbb{Z}_N)^4$ center symmetry without requiring the large-$N$ tuning of the twist parameter $k$. 
This implies that the TEK model with the minimal twist and heavy adjoint fermions exhibits large-$N$ equivalence to four-dimensional $SU(N)$ Yang-Mills theory in a broader range of the parameter space, opening new possibilities for computing physical quantities at large $N$.
See Figs.~\ref{fig:P1} and \ref{fig:Phase1} for these results.
In particular, the ability to define gauge-invariant observables with emergent momentum-space interpretation, originating in the nontrivial twist $k\not=0$, allows us to measure physical observables such as the meson spectrum and the correlation functions in this framework; Ref.~\cite{Gonzalez-Arroyo:2012euf} computed the meson spectrum and the chiral condensate using the TEK model, and our setup provides an alternative route to compute them. 
These directions will be pursued in future work.

By partially expanding the reduced model along one direction to include a compactified dimension, we investigated the behavior of the theory on $\mathbb{R}^3 \times S^1$. We found indications that with periodic adjoint fermions, the model exhibits no deconfinement transition as the circle size is reduced in the parameter region explored, which is a concrete realization of the adiabatic continuity conjecture, suggesting that the center-symmetric (confined) phase of large-$N$ adjoint QCD is smoothly connected between large and small spatial circles.
See Figs.~\ref{fig:AC1}, \ref{fig:AC2}, \ref{fig:PW} and \ref{fig:MT} for these results.
As a contrast, we confirmed that, when fermions are given antiperiodic (thermal) boundary conditions, the expected deconfinement transition occurs, indicating that the periodicity of fermions is crucial to the continuity. 
We also studied in detail the volume independence in the partially reduced TEK models both with the symmetric twist and the modified twist. 
We obtained the numerical results indicating the adiabatic continuity in the model with the modified twist, which is confirmed to preserve the volume independence.
Furthermore, we investigate the possible interplay between confinement and conformality in the small-mass region in QCD(adj.) with $N_{f}=2$.

In summary, the adjoint TEK model provides a versatile and efficient tool for probing large-$N$ dynamics of gauge theory. By leveraging heavy adjoint fermions to stabilize the center, one may be able to avoid tuning the twist parameter $k/L$ and directly simulate physics on a one-site lattice while keeping the minimal twist $k=1$. One also finds the adiabatic continuity of ${\mathbb Z}_N$ symmetric phase, or the absence of the deconfinement phase transition in the partially reduced adjoint TEK model with relatively light adjoint fermions.
Our results lend support to the theoretical conjectures about volume reduction and adiabatic continuity in gauge theories with adjoint matter.

For future investigations, we list several important topics:

\begin{itemize}

\item
It is obviously important to extend the parameter regions.
In the one-site adjoint TEK model, we will increase $b$ to show whether the $({\mathbb Z}_N)^4$ symmetric phase extends to $b=\infty$ or the continuum limit.
Furthermore, to calculate observables such as the meson mass spectrum in the adjoint TEK framework with heavy quark mass, we need to increase $N$ further.
In partially reduced adjoint TEK model, we consider to broaden the parameter region of ($\kappa,b$) to clarify the confined-deconfined phase structure.

\item
We will in detail study the modified twist in Eq.(\ref{eq:mtwist})  for the partially reduced model in larger $N$ although we gave results on adiabatic continuity with the twist for small $N$ in Sec.~\ref{sec:MT}.
We will compare the numerical results from the partially reduced models with the symmetric and modified twists for larger $N$, and will argue which is a better setup for the model on ${\mathbb R}^3 \times S^1$. 

\item
In semi-classics of continuum field theory, $SU(N)$ gauge theory with heavy adjoint fermions and $N$-flavor light fundamental fermions on ${\mathbb R}^3 \times S^1$ have been investigated \cite{Cherman:2017tey}, where ${\mathbb Z}_N$-twisted boundary condition is imposed on the fundamental fermions in $S^1$ \cite{Kouno:2012zz, Sakai:2012ika, Kouno:2013zr, Kouno:2013mma, Kouno:2015sja, Iritani:2015ara, Misumi:2015hfa, Shimizu:2017asf, Tanizaki:2017qhf, Tanizaki:2017mtm, Dunne:2018hog}. It is interesting to construct the TEK model and the partially reduced model corresponding to this setup.

\item
Finally, we comment on ongoing work about the $N_f = 1$ adjoint TEK model.
For $N_f = 2$ adjoint QCD, the theory may approach a conformal regime in the massless limit, thus it is not completely conclusive whether the continuity of the center-symmetric phase we observed is that of ``usual" confined phases, as discussed in Sec.\ref{sec:PDCC}.
To resolve this ambiguity, we are currently investigating the $N_f = 1$ adjoint TEK model. Since the theory with a single adjoint flavor is believed to be confining even in the chiral limit, this setup will allow a cleaner test of adiabatic continuity in a genuinely confining context \footnote{The adjoint QCD with $N_f=1/2$ (a single Weyl fermion) corresponds to ${\mathcal N} =1$ super Yang-Mills theory (SYM), where the adiabatic continuity of confined phase is believed to exist. The adjoint TEK model with $N_f=1/2$ has been studied \cite{Butti:2022sgy,Bonanno:2024bqg,Bonanno:2024onr}.}.


\end{itemize}

\begin{acknowledgments} 
The authors thank M.~Okawa, A.~Gonzalez-Arroyo and M.~Unsal for reading the draft and giving us useful comments.
The authors appreciate the YITP workshop ``Progress and Future of non-perturbative quantum field theory" (YITP-W-25-25) and the YITP long-term workshop ``Hadrons and Hadron Interactions in QCD 2024'' (YITP-T-24-02) for providing the opportunities of useful discussions.  This work was partially supported by Japan Society for the Promotion of Science (JSPS) KAKENHI Grant No.~23K03425 (T.M.) and 22H05118 (T.M.).
The numerical calculations were partly carried out on Yukawa-21 at YITP in Kyoto University.
\end{acknowledgments}


\begin{appendix}

\begin{table}[tbp]
\begin{tabular}{|p{1cm}|p{1cm}|p{1cm}|p{2cm}|p{2cm}|p{2cm}|p{2cm}|}
\hline
$N$                    & $b$                     & $k$                  & $\kappa$     & $\#$ of config. & $W(1, 1)$    & $P$          \\ \hline \hline
\multirow{1}{*}{121} & \multirow{1}{*}{0.30} & \multirow{1}{*}{1} & 0.01               & 50            & 0.312090   & 0.007093   \\ \hline
\multirow{3}{*}{121} & \multirow{3}{*}{0.35} & \multirow{3}{*}{1} & 0.005              & 50            & 0.496952   & 0.263992   \\ \cline{4-7}
                     &                       &                    & 0.01               & 50            & 0.517899   & 0.017501   \\ \cline{4-7}
                     &                       &                    & 0.02               & 100           & 0.524526   & 0.003107   \\ \cline{4-7} \hline
\multirow{4}{*}{121} & \multirow{4}{*}{0.36} & \multirow{4}{*}{0} & 0.005              & 100           & 0.698076   & 0.587818   \\ \cline{4-7} 
                     &                       &                    & 0.01               & 100           & 0.660069   & 0.478603   \\ \cline{4-7} 
                     &                       &                    & 0.02               & 100           & 0.603082   & 0.411000   \\ \cline{4-7} 
                     &                       &                    & 0.03               & 100           & 0.596505   & 0.122023   \\ \hline
\multirow{4}{*}{121} & \multirow{4}{*}{0.36} & \multirow{4}{*}{1} & 0                  & 100           & 0.525707   & 0.341664   \\ \cline{4-7} 
                     &                       &                    & 0.005              & 100           & 0.551046   & 0.009968   \\ \cline{4-7} 
                     &                       &                    & 0.01               & 100           & 0.555670   & 0.006961   \\ \cline{4-7} 
                     &                       &                    & 0.02               & 100           & 0.557402   & 0.002117   \\ \cline{4-7}
                     &                       &                    & 0.03               & 100           & 0.553372   & 0.001658   \\ \cline{4-7} \hline
\multirow{1}{*}{121} & \multirow{1}{*}{0.40} & \multirow{1}{*}{1} & 0.005              & 50            & 0.625206   & 0.002447   \\ \cline{4-7}
                     &                       &                    & 0.01               & 50            & 0.624670   & 0.005130   \\ \cline{4-7} \hline
\multirow{1}{*}{121} & \multirow{1}{*}{0.45} & \multirow{1}{*}{1} & 0.005              & 50            & 0.678856   & 0.003658   \\ \hline

\end{tabular}
\caption{Numerical values and simulation parameters used in Section \ref{sec:ZN} and the resulting Wilson loop $W(1,1)$ and Polyakov loop.}
\label{tab:SIM}
\end{table}

\section{Details of the simulations}
\label{sec:SS}
In this paper, we use the standard Hybrid Monte Carlo (HMC) algorithm. We invert $Q^2 (Q = D_W \gamma_5)$ using the Conjugate Gradient (CG) algorithm. Our stopping condition is as follows. Let $r = b - Q^2 x$ be the residue with $b$ being the source. Then we require $|r|^2 / |b|^2 < 10^{-7}$ during the molecular dynamics evolutions, and $|r|^2/|b|^2 < 10^{-15}$ at the global reject-accept step. The step size in the molecular dynamics evolution and the number of time steps in one trajectory were set so that the acceptance rate exceeded 50\%. In the simulation described in Section \ref{sec:ZN}, we generated 300 trajectories for each value of $\kappa$ with fixed $b$, and computed the expectation values of the Polyakov loop and Wilson loop using the last 100 trajectories. Only the simulations for $N = 121, b = 0.30, 0.35, 0.40, 0.45$ used the last 50 trajectories out of 150 trajectories.
For all simulations in Section \ref{sec:AC}, 300 trajectories were generated and the last 100 trajectories were used.
Most of simulations were performed using a desktop PC equipped with a 12th Gen Intel(R) COre(TM) i9-12900K 3.20GHz. 
Some of simulation settings and results are shown in Table \ref{tab:SIM}.



\end{appendix}

\clearpage

\bibliographystyle{utphys}
\bibliography{./QFT,./refs,./references}

\end{document}